# A Cryogenic Telescope for Far-Infrared Astrophysics: A Vision for NASA in the 2020 Decade

*A white paper submitted to NASA's Cosmic Origins Program Office*

C.M. Bradford *[1,2], P.F. Goldsmith[1], A. Bolatto[4], L. Armus[2,3], J. Bauer[3,1], P. Appleton[2], A. Cooray[5,2], C. Casey[5], D. Dale[6], B. Uzgil[7,2], J. Aguirre[7], J.D. Smith[8], K. Sheth[10], E.J. Murphy[3], C. McKenney[1,2], W. Holmes[1], M. Rizzo[9], E. Bergin[11] and G. Stacey[12]

[1]Jet Propulsion Laboratory
[2]California Institute of Technology
[3]Infrared Processing and Analysis Center, Caltech
[4]University of Maryland
[5]University of California, Irvine
[6]University of Wyoming
[7]University of Pennsylvania
[8]University of Toledo
[9]Goddard Space Flight Center
[10]NRAO, Charlottesville
[11]University of Michigan
[12]Cornell University

May 7, 2015

## Abstract

Many of the transformative processes in the Universe have taken place in regions obscured by dust, and are best studied with far-IR spectroscopy. We present the Cryogenic-Aperture Large Infrared-Submillimeter Telescope Observatory (CALISTO), a 5-meter class, space-borne telescope actively cooled to T$\sim$4 K, emphasizing moderate-resolution spectroscopy in the crucial 35 to 600 $\mu$m band. CALISTO will enable NASA and the world to study the rise of heavy elements in the Universe's first billion years, chart star formation and black hole growth in dust-obscured galaxies through cosmic time, and conduct a census of forming planetary systems in our region of the Galaxy. CALISTO will capitalize on rapid progress in both format and sensitivity of far-IR detectors. Arrays with a total count of a few $\times 10^5$ detector pixels will form the heart of a suite of imaging spectrometers in which each detector reaches the photon background limit.

The Far-IR Science Interest Group will meet from 3–5 June 2015[1] with the intention of reaching consensus on the architecture for the Far-IR Surveyor mission. This white paper describes one of the architectures to be considered by the community. One or more companion papers will describe alternative architectures.

*bradford@caltech.edu

[1]http://conference.ipac.caltech.edu/firsurveyor/

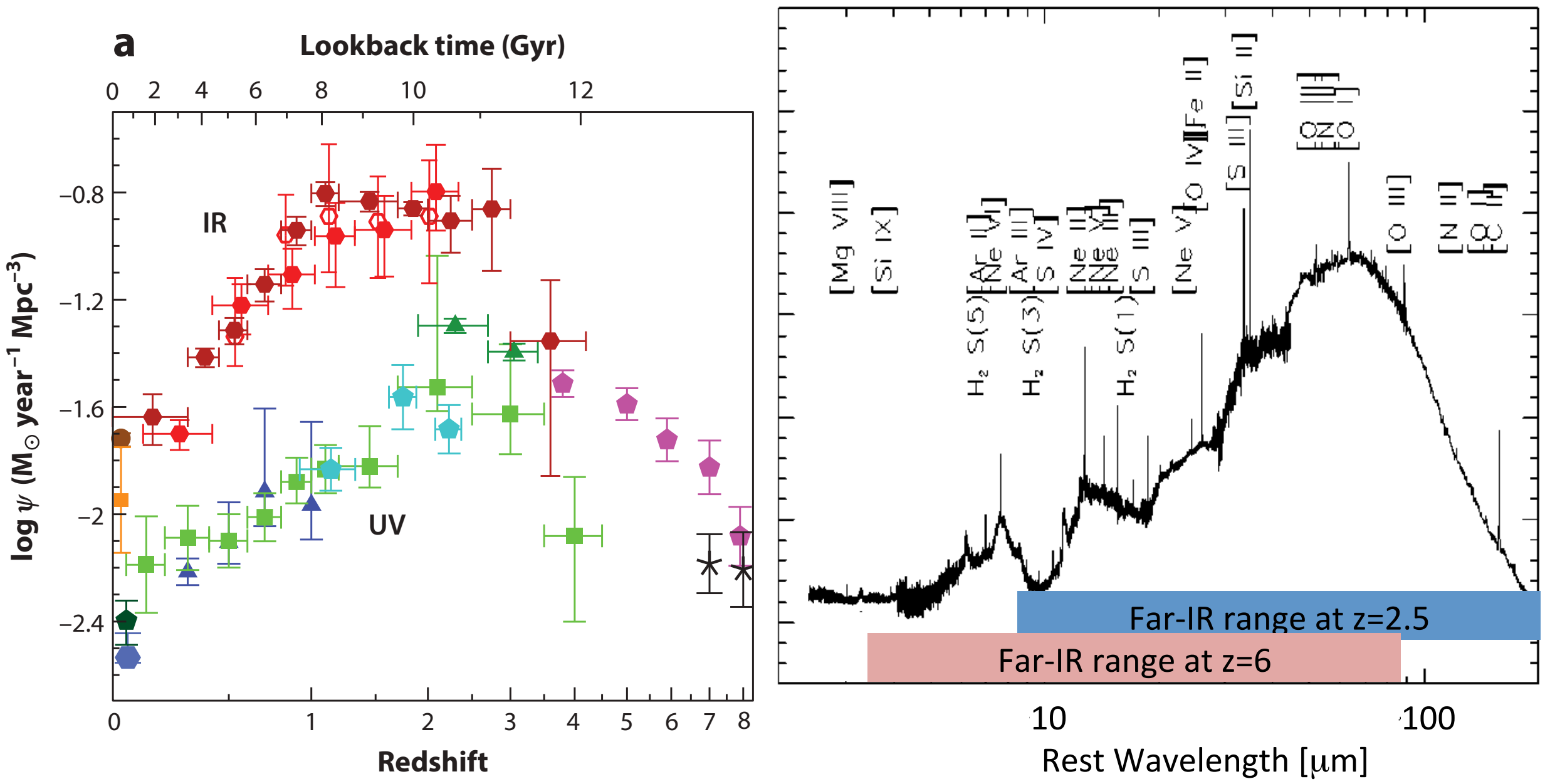

Figure 1: LEFT Cosmic star formation rate history as measured in the rest-frame ultraviolet, and far-infrared, reprinted from Madau & Dickinson, 2014 [44]. Red points are from Spitzer and Herschel, green and blue from rest-frame UV surveys. Purple points are from the Bouwens et al. [7, 6] based on deep Hubble fields using dropout selections. Right: Full-band spectrum of Circinus, a nearby galaxy with an active nucleus obscured by dust, obtained with the Infrared Space Observatory (ISO) [25, 50, 63]. This shows the range of ionized, atomic, and molecular gas cooling lines originating deep in the obscured core of the source. (Vertical axis is $\lambda F_\lambda$, major ticks $5\times10^{-12}\,\mathrm{W\,m^{-2}}$.)

# 1 Introduction, Motivation

After the Cosmic Microwave Background (CMB) is accounted for, the remaining cosmic background light is the integrated emission from all stars and galaxies through cosmic time. The spectrum of this cosmic background shows two broad peaks with comparable observed flux density, one at $\sim$1 $\mu$m, and one at $\sim$150 $\mu$m. The long-wavelength component, called the Cosmic Infrared Background (CIB) [23, 19], is radiation from dust heated by stars or accreting black holes. Its prominence is a simple consequence of the fundamental link between the star formation and its fuel: the interstellar gas with obscuring dust. We now have strong evidence that most of the energy that has been produced by galaxies through cosmic time has emerged in the far-IR [54, 44]. The typical UV/optical photon from a young star has been absorbed by dust and re-radiated (see Figure 1).In general, rest-frame ultraviolet and optical-wavelength light does not access the obscured regions that dominate the activity in galaxies. Similarly, in nearby galaxies and in our own Milky Way, star-forming cores, embedded young stars, and protoplanetary disks all cool primarily through the far-infrared.

The spaceborne Spitzer and Herschel observatories have demonstrated the importance of the far-IR waveband, but it is only with sensitive spectroscopic capability that astronomers will have the opportunity to study in detail the inner workings of galaxies at cosmological distances and late-stage forming planetary systems. This capability has not yet been realized because it requires a combination of a cold telescope and very sensitive direct detectors. After 2 decades of development of superconducting detectors, and with the system-level experience gained with previous-generation cryogenic satellites, we are now in a position to field CALISTO, a large space telescope actively cooled to a few degrees K with $\sim 10^5$ individual far-IR detector pixels, each operating at the fundamental sensitivity limit set by the astrophysical background. This paper builds on the concept for CALISTO presented late last decade [26, 10]; it will be a large facility-class observatory launched to an earth-sun L2 halo orbit with at least a 5-year design lifetime. The key advance in the last decade is the progress in far-IR detector technology.

# 2 Motivation for Sensitive Wideband Far-IR Spectroscopy

The sensitive CALISTO platform is especially compelling for wideband spectroscopy, as Figure 2 shows and Table 1 presents. CALISTO will obtain full-band spectra of thousands of objects ranging from the first dusty galaxies to the most heavily enshrouded young stars and proto- planetary disks in our own Galaxy, as well as blind discovery of thousands more. These CALISTO spectra will directly address several key goals of modern astrophysics:

2.1 Measure the onset of heavy elements and the rise of organic molecules in the Reionization Epoch.
2.2 Chart the true history of cosmic star formation and its connection to supermassive black hole growth.
2.3 Measure clustering and total emission of faint galaxies below the individual detection threshold using tomographic intensity mapping of the far-IR emission lines.
2.4 Probe the cycling of matter and energy in the Milky Way and nearby galaxies.
2.5 Conduct a census of gas mass and conditions in protoplanetary disks throughout their evolutionary sequence.
2.6 Assess the origin, transport and cooling role of water in sources ranging from the solar system to distant galaxies.

## 2.1 The Reionization Epoch: the Rise of Heavy Elements and Dust

As the Universe is enriched from primordial $H_2$ to a medium which contains heavy elements and dust grains, the key cooling pathways shift from the quadrupole pure rotational $H_2$ lines (28, 17, 12, 9.7, 8.0, 6.9... $\mu$m) to a combination of the fine-structure transitions and the dust. CALISTO will probe all phases of this transition. For metallicity above $\sim 10^{-4}$ solar, fine-structure lines are believed to become more important than $H_2$ for gas cooling [55]. However, surprisingly powerful $H_2$ emitters (e.g., Stephan's Quintet, Zw3146, and the z=2.16 'Spiderweb' protocluster) have been found at low-redshift with Spitzer [51, 1, 16, 21, 53, 52]. Sources like Stephan's Quintet may be analogs of early-Universe shocks produced in galaxy formation and AGN feedback, when dust and metals are emerging from the first cycles of enrichment. *For z $\sim$5–10, the $H_2$ lines are redshifted into the far-IR, and remarkably, sources like Zw3146 and the 'Spiderweb' would be detectable in their $H_2$ lines with CALISTO even at z $\sim$8–10.*

Once heavy elements are in place, the rest-frame mid-IR dust features may actually be the most practical probe of heavy elements at early times due to their large equivalent widths. Dust is believed to form as the first heavy elements are created, for example in pair-instability Population III supernovae remnants, and Spitzer has shown that the dust features are often the brightest features in the spectra of galaxies at all wavelengths. In particular, the polycyclic aromatic hydrocarbon (PAH) features at 6.2–17 $\mu$m are unambiguous, with to 15$\times$ more power than the brightest atomic cooling lines, and act as sensitive probes of heavy element abundance for metallicity $<$0.2 [22]. Like the $H_2$ lines, most features are redshifted out of the JWST band, but not into the ALMA windows in the z $\sim$5–10 era. CALISTO can detect these powerful bands at early epochs (Fig. 2), thus probing the transition from primordial $H_2$ to heavy-element cooling in the Universe's first Gyr.

We refer the interested reader to white papers by Appleton et al., and Cooray et al. for further discussion on these aspects.

## 2.2 Charting the Cosmic History of Star Formation and Black Hole Growth

Far-IR and submillimeter continuum imaging surveys are now revealing cosmologically-significant populations of high-redshift galaxies which are so highly obscured that they emit nearly all of their energy in the mid-IR through submillimeter. These datasets, as well deep X-ray surveys show that much of the formative growth of stellar populations and black holes has been deeply obscured by dust for the bulk of the Universe's history, and thus inaccessible to astronomers' traditional diagnostic toolkit: rest-frame optical spectroscopy. With its excellent spectral sensitivity in the 35–600 $\mu$m band, CALISTO brings a powerful new toolkit to bear on these high-redshift galaxy populations: the rest-frame mid- to far-IR, where the dust becomes optically thin, and the dominant interstellar coolants lie (Fig. 1, right). CALISTO spectra of distant galaxies will:

- Provide an unambiguous redshift, or look-back time for each galaxy.
- For each, determine the total star formation rate in the galaxy and infer a spatial scale of the buried starburst regions [61] by comparing the intensities of the atomic gas coolants—$Si^+$, $C^+$, and $O^0$—with the total far-IR continuum intensity. (The star formation extent may or may not be related to the spatial extent of the molecular gas reservoir which will be directly imaged with ALMA.) In aggregate, these measurements chart the time history of dust-obscured stellar power output, .
- Estimate the top end of the stellar mass function via its effect on the UV field and the resulting ionization structure reflected in the fine-structure lines of ions: $O^{++}$, $Ne^{++}$, $N^{++}$, $S^{++}$, and $N^+$, $Ne^+$.

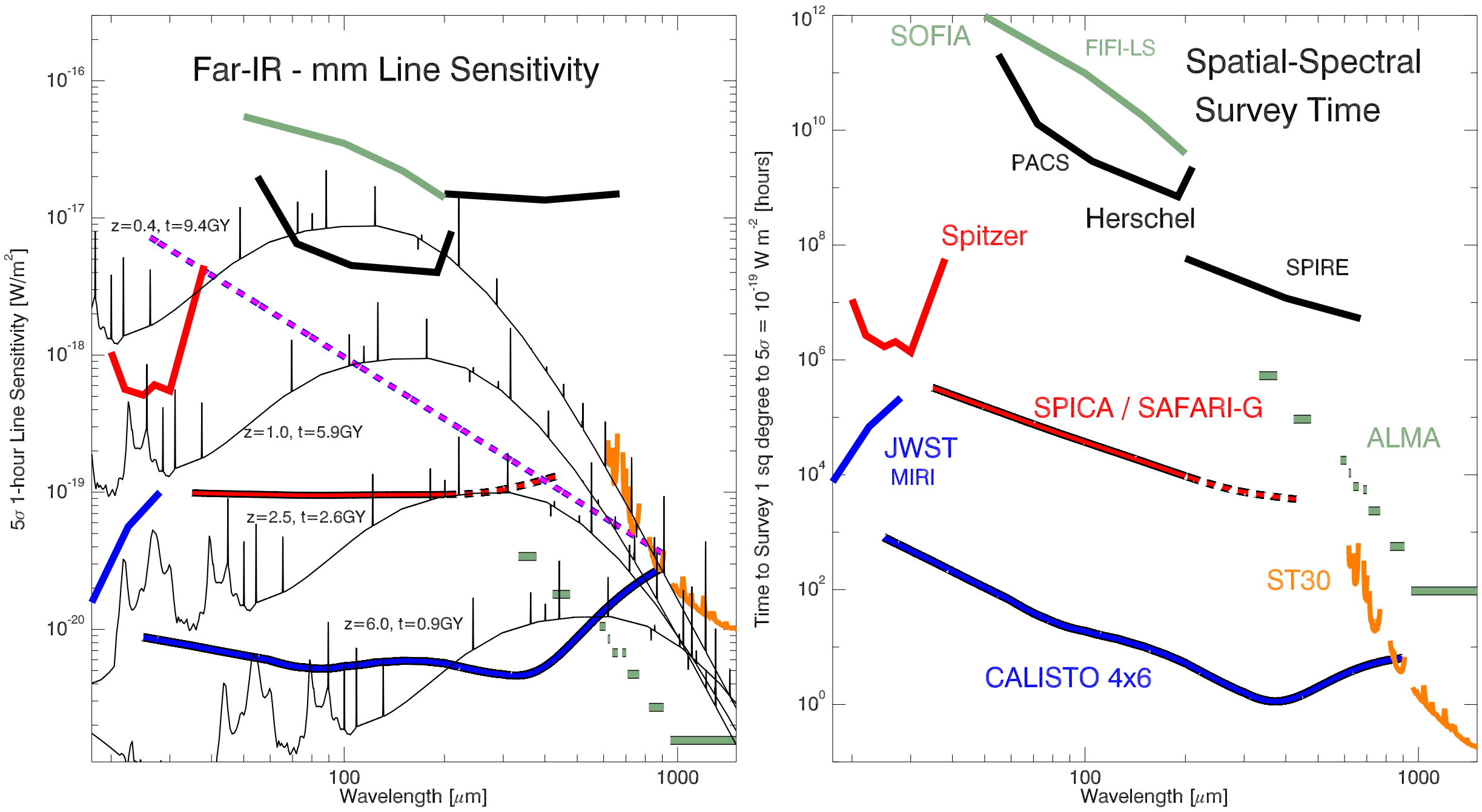

Figure 2: Spectroscopic sensitivities in the far-IR and submillimeter. Left shows the sensitivity in $\mathrm{W\,m^{-2}}$ for a single pointed observation. Galaxy spectra assuming $\mathrm{L} = 10^{12}\,\mathrm{L}_\odot$ at various redshifts are overplotted using light curves, with continuum smoothed to R=500. The magenta dashed line shows the sensitivity of a quantum-limited heterodyne receiver ($T_{sys}$=h$\nu$/k) in a bandwidth of 10 km/s. The right panel shows the speed for a blind spatial-spectral survey reaching a depth of $10^{-19}\,\mathrm{W\,m^{-2}}$ over a square degree, including the number of spatial beams and the instantaneous bandwidth. CALISTO 4×6 refers to the baseline configuration, assuming R=500 grating spectrometers with 100 beams (a conservative figure) and 1:1.5 instantaneous bandwidth. Detectors are assumed to operate with NEP = $2\times10^{-20}$ $\mathrm{WHz^{-1/2}}$, a figure which has been demonstrated in the lab. The SPICA / SAFARI-G curve refers to the new configuration: a 2.5-meter telescope with a suite of R=300 grating spectrometer modules with 4 spatial beams, and detectors with NEP=$2\times10^{-19}$ $\mathrm{WHz^{-1/2}}$. ST30 represents a 30-meter class wide-field submillimeter telescope in the Atacama, such as CCAT, equipped with 100 spectrometer beams, each with 1:1.5 bandwidth. ALMA band averaged sensitivity, and survey speed based on 16 GHz in the primary beam.

- Where present, directly measure the highly-ionized gas around the AGN itself with fine-structure transitions of high-ionization-state species such as $Ne^{4+}$ and $O^{3+}$ (ionization potential of 97 & 54 eV, respectively).
- Probe the warm (∼1000 K), dense ($10^7$ $\mathrm{cm^{-3}}$) molecular torus believed to exist around AGN. —a likely way-point as material is funneled from the host galaxy down to the accretion zone. It is expected to emit strongly in the high-J CO rotational transitions ($\lambda_{rest}$ ∼50–80 $\mu$m), easily detectable with CALISTO to z=5.
- In aggregate thereby track the fraction of energy release due to accretion and its relationship to the star-formation history.

Further information can be found in the Armus et al., whitepaper.

We emphasize that the excellent sensitivity of CALISTO is essential for these distant-galaxy measurements. Charting a complete history requires study of galaxies before, during, and since the putative era of peak star formation and black hole growth 2–6 Gyr after the Big Bang. To reach the first Gyr of the Universe (z=6) in the spectral probes demands a line sensitivity below $10^{-20}\,\mathrm{W\,m^{-2}}$, which in the far-IR is only achieved with an actively-cooled telescope and an optimized dispersive spectrometer as baselined for CALISTO. CALISTO will be used in 2 ways. 1) The instantaneous wideband coverage permits rapid follow-up of individual sources of interest, discovered for example with ground-based submillimeter continuum surveys, LSST, JWST or Euclid. 2) The simultaneous spatial multiplexing enables blind spatial / spatial surveys, which will discover many line-emitting sources blindly (on order 3–30 per 1-hour pointing) as well as reveal the underlying 3-D clustering of undetected sources in the residual signal.

Table 1: CALISTO Spectrometer Backends: R=500 Strawman Design

| Parameter | 40 $\mu$m | 120 $\mu$m | 400 $\mu$m | Scaling w/ $D_{\rm eff}$ |
|---|---|---|---|---|
| Dominant background | zodi dust | zodi. + gal. dust | tel. + CMB | ... |
| Photon-noise limited NEP [$\rm W\,Hz^{-1/2}$] | 3e-20 | 3e-20 | 4e-20 | ... |
| Beam size | 1.9$''$ | 5.9$''$ | 19$''$ | $\propto D^{-1}$ |
| Instantaneous FOV [sq deg] | 4.0e-5 | 3.8e-4 | 2.3e-3 | $\propto D^{-2}$ |
| Line sensitivity $\rm W\,m^{-2}$, $5\sigma$, 1h | 4.2e-21 | 3.3e-21 | 3.2e-21 | $\propto D^{-2}$ |
| Pt. sce. mapping speed [$\rm deg^2/(10^{-19} W\,m^{-2})^2/sec$] | 1.6e-4 | 2.4e-3 | 1.6e-2 | $\propto D^2$ |
| Surface bright. sens. per pix [MJy/sr $\sqrt{\rm sec}$] | 4.2 | 1.1 | 0.33 | $\propto D^0$ |

Notes: Sensitivities assume single-polarization instruments with a product of cold transmission and detector efficiency of 0.25 in a single polarization, and an aperture efficiency of 0.75. FOV estimate assume slit widths of 165 $\lambda/D$ for the 40 and 120 $\mu$m examples, and 100 individual single-beam spectrometer backends for the 400 $\mu$m case.

### 2.3 Tomographic Intensity Mapping: Measuring Clustering and Absolute Cosmic Line Intensities.

In addition to studying individual galaxies, the large-throughput wide-band spectrometers on CALISTO will carry out blind 3-D intensity mapping, or tomography, using the bright far-IR fine-structure transitions described above (especially NeII, OI, OIII and (for $z < 2$) CII). This is an emerging technique that has grown out of the 21-cm tomography experiments and instruments targeting early-Universe CO and CII from the ground are now under development. As outlined in Visbal & Loeb 2010 [68, 69], Gong et al., 2011, 2012, 2013 [29, 28, 27], and Uzgil et al., 2014 [67] these datasets will reveal 3-D clustering due to large scale structure, even when individual galaxies are not detected. The amplitude of the clustering signal is the product of the galaxy to dark matter bias and the total mean intensity of a given spectral feature. With reasonable assumptions about the bias, this measurement can then can thus probe the total cosmic luminosity of each of the fine-structure transitions as a function of time, with a built-in redshift precision not available to the continuum surveys. It is a promising approach both for assessing the contributions of faint galaxies, particularly important early in the Universe's history. For good sensitivity to the large-scale signal, mapping over $\sim$1–2 square degrees, with large depth in redshift provided by the spectrometer is sufficient; as Table 1 shows, this is possible for CALISTO with the high-throughput spectrometer.

### 2.4 Galaxy Archeology and Cycling of Matter in the Milky Way and Nearby Galaxies

Observations of the distant Universe are, by necessity, interpreted in the context of the Milky Way and nearby galaxies. These provide the windows into the details of the astrophysical processes that drive galaxy evolution: cycling of matter between stars and the interstellar medium (ISM), self-regulation of star formation, formation of stars on galaxy scales, and feedback from central AGN. With its exquisite surface brightness sensitivity (Table 1, 5th row), CALISTO will provide unparalleled mapping speed for integrated line and continuum emission with useful angular resolution (e.g. 8$''$ at 158 $\mu$m gives 300 pc resolution at 10 Mpc, or 1 kpc resolution at 25 Mpc). As noted above, the 35 to 600 $\mu$m region of the spectrum has a number of key transitions for the cooling of the neutral and molecular gas and the probing of ionized material: namely the bright fine structure transitions of [CII], [OI], [NII], [OIII], possibly [CI] (depending on the long-wavelength cutoff) and CO and $H_2O$ among others.

PACS on Herschel gave us a flavor for the type of science that these observations enable. In particular, studies have shed considerable light on how and where emission from [CII] is produced in the Milky Way and how and why it is a useful tracer of star formation [**? ?** ]. But these observations have been limited by the low mapping speed and sensitivity, yielding only very sparse samples.

The sensitivity of CALISTO will be a huge leap forward. The angular resolution of 8$''$ corresponds to 0.16 pc at 5 kpc, so can distinguish diffuse regions from cloud surfaces at this distance. But the biggest advance will be in sensitivity and mapping speed. The surface brightness sensitivity is independent of beamsize (thus telescope aperture) and can be translated directly into column density sensitivity for a given species if the gas excitation known (e.g. Crawford et al., 1985 [17], Madden et al., 1997 [45]). For example, for [CII] in atomic gas at T = 100 K, $\rm n_H = 10\,cm^{-3}$ (so [CII] is sub-thermally excited), CALISTO can detect ($5\sigma$) a column of $\rm N_H = 1.8\times10^{19}\,cm^{-2}$ in 100 sec. A similar result is obtained for ionized gas with an electron density of only 0.05 cm$^{-3}$ (so [CII] is again sub-thermal). Denser gas is of course much easier to detect per unit column density. This sensitivity is multiplexed both spatially ($\sim$100 beams) and spectrally (full-band coverage), so that by rastering CALISTO's multi-beam spectrographs, it will be possible to map large regions in the Galaxy, and *thousands* of nearby galaxies in key far-IR transitions. The speed could be increased further if only one or a few individual lines are desired by using a dedicated Fabry-Perot type spectrograph coupling 2 spatial dimensions (so $\sim$several thousand beams). While 2-D datasets could be obtained fairly quickly with the direct detection spectrometers, we emphasize that a heterodyne receiver array would provide velocity resolved images of clouds, enabling disentangle the structures along the line of sight, and developing full kinematic picture of the ISM.

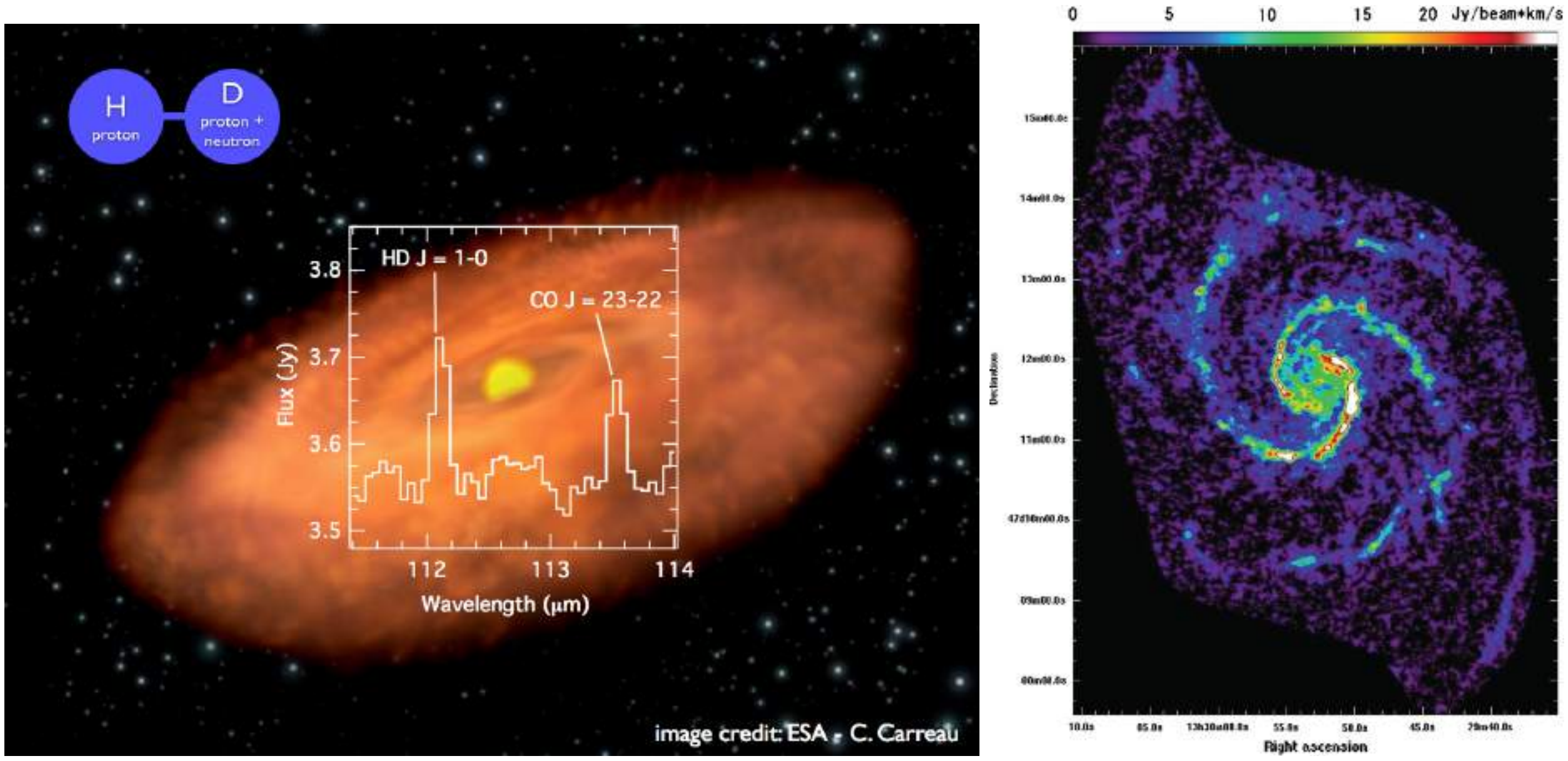

Figure 3: LEFT: Hydrogen deuteride (HD) detected in the TW Hya protoplaentary disk, superposed on an artists conception of a young gas-rich disk. While not the strongest feature in the spectrum, HD is an excellent tracer of total molecular hydrogen mass, and CALISTO will be able survey HD as well as other key coolants in hundreds to thousands of such systems at various evolutionary stages reaching kilo-parsec distances. RIGHT: CO $\mathrm{J}=1\rightarrow 0$ map in M51 obtained with CARMA, reprinted from Koda et al. [38]. CALISTO will offer only $2\times$ coarser resolution in [CII] with the speed to make large maps, providing a full census of the ISM phases in this and other nearby spiral galaxies.

For galaxies, the resolution will be comparable to what the VLA obtains for HI, though column density sensitivity is much better than the VLA at this resolution; it is a better match to the anticipated sensitivity of the SKA on this sizescale. These measurements will provide a broad range of physical information such as metallicity indicators, radiation field estimators (from dust and line emission), and gas heating measurements. The maps will also reveal galaxy-scale galactic outflows in ionized species such as [CII], as well as faint outer disks and extra-planar structures.

These maps will be vital for understanding the star formation process in spiral galaxies. As an example, [CII] mapping in M51 will complete our tracking of the ISM phases, probing both the atomic PDR gas and the CO dark $H_2$ gas over the entire disk, including both arm and interarm regions. Figure 3 (right) shows an image of M51 in the lowest rotational transition of the carbon monoxide molecule, which traces the purely molecular portion of the interstellar medium [38]. A comparable [CII] map with CALISTO will enable a more complete understanding of both the interactions with galactic spiral structure and star formation activity. The $8''$ angular resolution of CALISTO will cleanly separate arm and interarm regions in M51 (and other nearby galaxies). We expect that the [CII] emission arises predominantly from PDRs and surrounding molecular clouds in the inter-arm and upstream side of spiral arms, while the primary source would change to photo-dissociated atomic gas and HII regions on the downstream side. Probing this phase transition will test both the density-wave theory, and provide a critical understanding for interpreting the [CII] emission in diverse galactic environments. Because the expected velocity width of the M51 [CII] emission within the CALISTO beam is only 5 to 20 km $\mathrm{s}^{-1}$, the line-to-continuum ratio in the baseline CALISTO grating spectrometers may be a concern for these measurements, and should be studied carefully. Of course, a heterodyne spectrometer is well-suited to the narrow lines. Furthermore, only with a heterodyne receiver can the 3rd dimension be provided – this enables discriminating gas on the upstream side from that on the downstream sides of the arms using their velocities ($\delta v \sim 30 - 60\,\mathrm{km/s}$).

## 2.5 Planetary-System Formation in the Milky Way: Gas in Disks

The evolution of circumstellar disks and their **gas** component is key to planet formation. Disks rapidly evolve from the primordial gas-rich phase to planetary systems largely devoid of gas. Even small amounts of residual gas at late stages can affect the settling and radial drift of dust grains, planetary migration, and eccentricity evolution. It is thus crucial important for understanding the formation of both terrestrial and Jovian worlds [65, 70, 39, 36]. Spitzer has detected many atomic, ionic and molecular gas emission lines that arise from the inner 1–20 AU regions of disks, but as Herschel has shown, the bulk of the disk mass is in the outer disk that emits primarily in the far-IR. The various emission lines in the CALISTO band originate from different regions of the disk and will trace the gas properties of disks as they evolve, form planets and eventually dissipate. In particular, the rotational fundamental of HD ($\lambda$=112 $\mu$m) has recently been shown using the Herschel PACS spectrometer to be a robust tracer for total gas mass in the closest planet-forming disk system [4]. HD is a direct analog to $H_2$ with similar chemistry and, with a small dipole moment, is emissive at the characteristic temperatures of the main disk mass reservoir.

Greater sensitivity is crucial for a full census of the evolutionary phases. Herschel was limited to the local (d$<$150 pc) star-forming environments where only low-mass star formation occurs. Only a small HD survey has been undertaken, producing 3 significant ($>3\sigma$) detections out of 7 systems. CALISTO will be approximately 1000 times more sensitive than Herschel PACS, thus probing the same source at $\sim$30$\times$ larger distances (so for example $>$1 kpc for the TW Hya system shown in Figure 3), in total some 30,000$\times$ larger volume. CALISTO will thus have access to many dense young clusters in giant molecular clouds, the dominant sites for low and high-mass star (and likely planet) formation.

## 2.6 Water in the Cosmos

Water is important both for its obvious astrobiological significance and because it is a critical coolant of star-forming gas. Recent Herschel observations have revealed water to be present in an enormous variety of regions in the solar system including asteroids (Ceres [40]), satellites (Enceladus [33]), comets, and planetary atmospheres (Jupiter [13]). The Herschel measurement of a deuterated to normal water abundance in comet 103P/Hartley2 identical to that on Earth [32], in comet 45P/Honda consistent with the Earth's value [43], but a much higher value measured by Rosetta in comet 67P/ChuryumovGerasimenko [42] has dramatically renewed interest in the role of cometary impacts for the origin of the Earths oceans. The only way to make progress in this important area is to observe a significant statistical sample of comets of different types, as well as other primitive bodies in the solar system. This will require very high spectral resolution, and heterodyne instrumentation is optimal. In particular, a modest heterodyne focal plane array covering the frequencies of appropriate $H_2^{18}O$ and HDO lines will be valuable as closer comets will be extended objects in CALISTO's beam.

In the Galaxy, water has been studied in a variety of interstellar regions by SWAS and Odin, with the general conclusion that it's abundance is low in the ISM. The much higher angular resolution and sensitivity of Herschel's HIFI instrument has shown that water can be a uniquely powerful tracer of the collapse of dense cores [37]. Extension of this work to even higher angular resolution and sensitivity should enable determination of the full three-dimensional velocity field in a star-forming core. Very high spectral resolution is optimal for this work—Herschel for example detected a *single* protostellar disk (TW Hydrae)in water, but with a line width of only 1.5 km/s [34]. A small heterodyne array operating at the frequencies of one (or more) of the lower water transitions is thus the ideal instrument here as well. The result would be a major, fundamental advance in understanding how stars and planets are formed. With higher sensitivity it should be possible to survey many nearby disks and determine their gas-phase water content.

Finally, we now know that water is the second-strongest molecular line emitter in nearby galaxies [71]. Existing data are all unresolved spectrally, but indicate the importance of water vapor as a tracer of shocks and as a coolant of dense gas. The ground state (557 GHz) water line has not been observed (the atmosphere is opaque even from the Atacama for redshift less than 0.02, or 6000 km/s) but is expected to be very intense. With beamsizes of a few arcseconds, the CALISTO spectrometers can map the water emission from nearby galaxies with sufficient sensitivity and angular resolution to probe of their spiral density wave structure, shocks, and star formation.

# 3 Wide-Field Continuum Mapping

While not the primary thrust of this paper, CALISTO is also very sensitive platform for continuum mapping, particularly at the short wavelengths where the confusion limit can be fairly deep. The deep confusion limit combined with the speed provided by low background platform enables surveys of large areas of sky to interesting depths. Table 2

Table 2: CALISTO Approximate Confusion Limits and Mapping Speeds

| $\lambda$ $\mu$m | Herschel $\sigma_C$ mJy | estimated $\sigma_C$ mJy | $\nu L_\nu$ z=2 $L_\odot$ | $\nu L_\nu$ z=5 $L_\odot$ | $NEF_{inst}$ mJy$\sqrt{s}$ | 5×time s | 5×time per sq deg h |
|---|---|---|---|---|---|---|---|
| 50 $\mu$m | 0.016 | 0.004 | 2.9e9 | 2.6e10 | 0.015 | 70 | 15 |
| 100 $\mu$m | 0.15 | 0.038 | 1.3e10 | 1.2e11 | 0.024 | 2.1 | 0.11 |
| 200 $\mu$m | 1.39 | 0.35 | 6.1e10 | 5.5e11 | 0.051 | 0.11 | 1.4e-3 |

Notes: Herschel $\sigma_C$ values are based on a power law implied by the 100 and 160 $\mu$m map RMS values in PACS deep fields (Magnelli et al., 2013 [46]). We simply reduce this by a factor of 4 to obtain an estimated $\sigma_C$ for CALISTO. Luminosity densities are then provided for 5× this depth, for z=2 and z=5. $NEF_{inst}$ is the raw instrument sensitivity. Times to confusion limit are conservatively estimated at 5× the time required for the instrument per-beam RMS to equal $\sigma_C$. The time to a square degree assumes a 4000-beam camera.

shows the estimated confusion RMS, obtained by simply reducing the Herschel PACS measured confusion limit (Magnelli et al., 2013 [46])by a factor of 4. This is a conservative since at these fluxes, the counts are becoming shallow, allowing the depth to be increased quickly with reduced beamsize. Some estimates suggest that the 100 $\mu$m confusion limit is 10× deeper at 5-m than at 3.5 m (see the white paper by Caitlin Casey et al. in this submission). The last two columns in Table 2 show 5× the integration time to reach this estimated confusion RMS, first per beam, and then per square degree, assuming a modest 4000-beam camera. The factor of 5 insures ample margin in the time estimate, and assures that instrument noise is sub-dominant to confusion. At 100 $\mu$m, the 38 $\mu$Jy depth corresponds to a Milky-Way type galaxy at z=2, well below the knee in the luminosity function for the peak of SF activity; this means that the bulk of the light is thus resolved into sources. The speed in this band is impressive; a full sky survey at 100 $\mu$m looks to be within reach in a $\sim$ 4000 hour survey with the strawman 4000-beam camera.

For nearby galaxies and the Milky Way, the continuum sensitivity at the short far-IR wavelengths is a powerful probe of tiny amounts of interstellar dust, complementing the gas-phase disk and ISM studies described above. At the distance of the Magellenic Clouds, for example, the sensitivity translates to $10^{-3}$ earth masses of dust. This opens the possibility to to carry out an essentially *complete survey of extragalactic debris disks around solar-type stars in the Clouds.*

### 3.1 Origin and Evolution of the Solar System studied with Trans-Neptunian Objects

Finally, we highlight a unique capability that CALISTO imaging provides for study of our Solar System's origins. The majority of small bodies in the solar system reside between 30 and 50 AU and are referred to as the Trans Neptunian Objects, or TNOs. These minor planets represent material from the origin of the solar system, unmodified by its subsequent evolution. They are the source of the short-period comets which deliver volatile materials to the inner solar system [3]. TNO orbital inclinations can be impacted by resonances with Neptune, and a census of TNO positions and orbital motions out to 100 AU provides information about the dynamical history of the outer solar system. These measurements have been difficult with optical-wavelength detection techniques, as the albedos can be small. With its excellent sensitivity in the deep thermal IR (e.g. $\sim$100 $\mu$m), CALISTO can probe the thermal emission of TNOs directly, reaching for example 140 km objects at 100 AU, deeper than existing optical surveys. Increased depth should be possible by looking for objects which move from observation to observation to observation in a given field; this should overcome the confusion limit. A second aspect to consider is the ability to detect halos of dust or possibly gas, some theories point to sublimation driven by CO even at several tens of AU [48]. Any such measurements of early cometary activity would constrain mass-loss rates, and the abundance of rarely-observed extremely-volatile species that may be relatively depleted in short-period comets

## 4 Architecture Choice

The scientific goals outlined above require excellent spectroscopic sensitivity, both for point sources and mapping, with full coverage between the 28 $\mu$m cutoff of JWST MIRI and the onset of the ground-based windows at $\sim$600 $\mu$m. Accessing the earliest galaxies and most-evolved lowest-mass protoplanetary systems requires a line sensitivity of $10^{-20}\,\mathrm{W\,m^{-2}}$, large instantaneous bandwidth, and moderate spectral resolving power ($R = \delta\lambda/\lambda \geq 500$). The requirement for ultimate sensitivity demands maximum collecting area, low telescope background, and high efficiency. Blind spectroscopic surveys over large fields will also be a part of the program, so the observatory must have enough throughput (A$\Omega$) to make use the large-format array technology now available. These crucial attributes are summarized in Table 3.

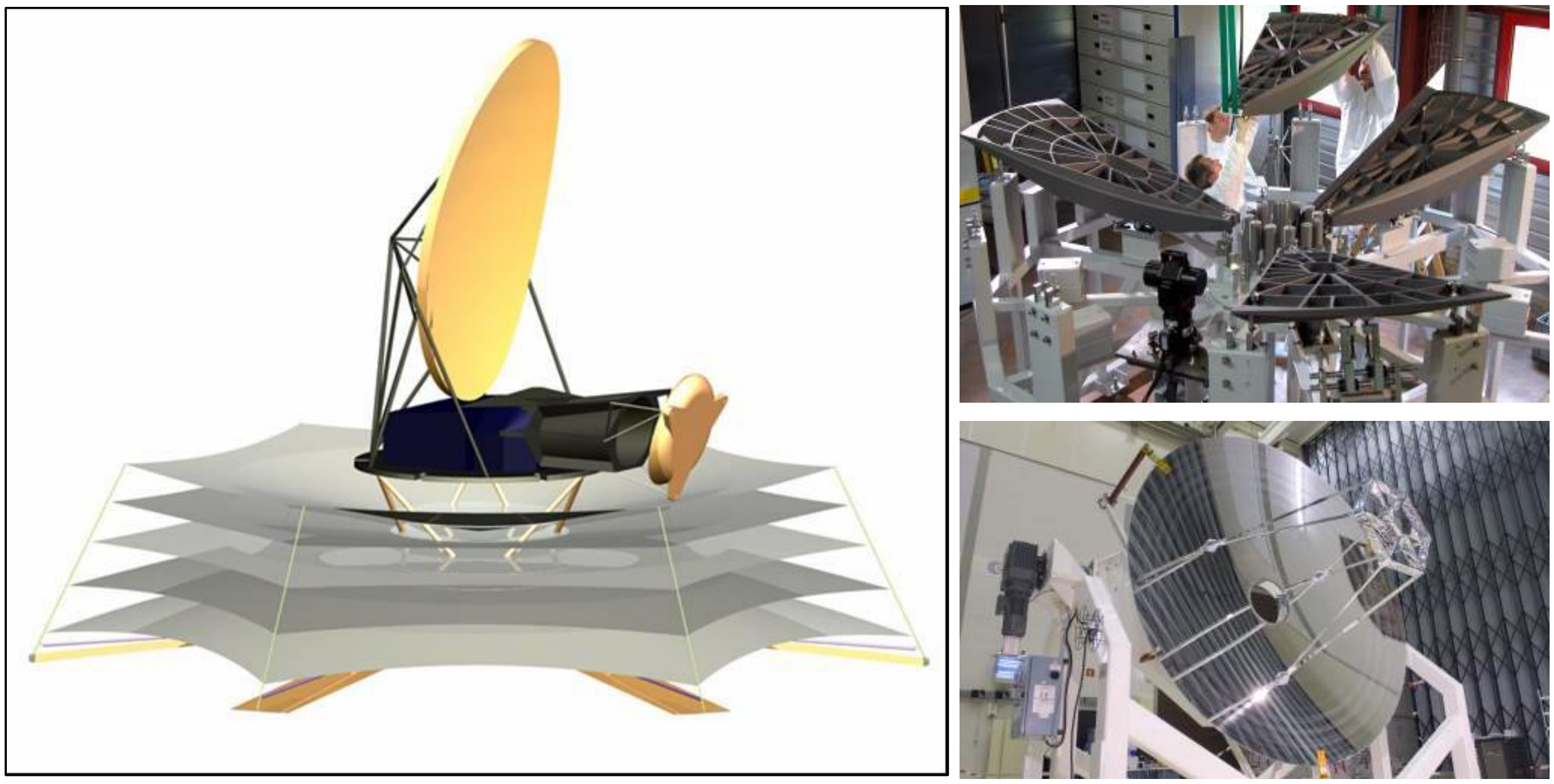

Figure 4: Left: CALISTO concept. 5-meter class telescope is actively cooled with closed-cycle coolers to ∼4 K. Passive and active cooling are integrated in a design which features V-groove radiators as used on Planck and JWST. Right: Large cold telescope heritage: the 3.5-meter Herschel silicon carbide primary mirror, prior to assembly from 8 petals and figuring, and as integrated into the telescope.

Collecting area per unit cost is maximized with a monolithic-aperture telescope, particularly since the entire telescope and instruments will be actively cooled. A single-dish telescope also naturally accommodates a wide range of instruments, for example the broadband imaging arrays, heterodyne receiver arrays, 2-D imaging spectrometers such as Fabry-Perot interferometers.

Table 3: CALISTO Basic Parameters

| Parameter | Value |
|---|---|
| Telescope Temperature | $<$4 K |
| Telescope Diameter | $\sim$5 m |
| Telescope Surface Accuracy | 1 $\mu$m |
| Telescope Field of View | 1 deg at 500 $\mu$m |
| Instrument Temperature | 50–100 mK |
| Total Number of Detectors | 1–5$\times 10^5$ |
| Heat Lift at 4 K | $\sim$150 mW |
| Heat Lift at 20 K | $\sim$2 W |
| Data Rate | tbd |

# 5 Observatory and Telescope

## 5.1 Observatory Cryogenics

Cooling all parts of the telescope and instrument environment to a few degrees K is essential for the excellent sensitivity, and this is a firm requirement for CALISTO. Cooling will be provided by closed-cycle helium coolers, carefully integrated into a passive cooling architecture which uses staged V-groove radiators. The effectiveness of the V-groove system has been demonstrated with the ESA Planck telescope, which reached below 40 K on orbit. 4-K class space-flight coolers have been developed by industries worldwide: in the US as part of NASA's ACDTP program, and in Japan by Sumitomo. The Sumitomo 4.5 K coolers have successfully flown and are now undergoing life-cycle tests in preparation for SPICA. A detailed thermal design will be part of the pre-decadal study, but the basic approach appears feasible. On aspect already clear is that the structure which supports the telescope for launch can not form the thermal path once on orbit, so a breakaway truss will be required. With this assumption, a conservative strawman estimate suggests that 150 mW of heat lift at 4 K will be ample, split roughly equally between overcoming the parasitics loads to the 4-K observatory, and supporting the sub-Kelvin coolers in the instruments. The most efficient design also employs active cooling at a state intermediate between the passive cooling floor and the cold telescope (e.g. 18–20 K), to the tune of $\sim$0.5 W for the parasitics, and perhaps another 1 W for the first stage amplifiers (see below). This can be naturally provided by Stirling stages in the US-built coolers, or as additional stand-alone coolers as in the Sumitomo architecture. The Sumitomo coolers require 2500 W supplied at the bus side per W of 4.5 K lift, or 450 W per watt at 18 K, so the total power requirement is $\sim$1500 W, including a factor of 1.5 margin for cooler degradation through the mission life.

The system will also likely use a set of dedicated 2-K class coolers to back the sub-K cooling for the instruments. Sumitomo has demonstrated such systems in preparation for SPICA; they are essentially the same as the 4-K systems, but they use $^3$He as the working fluid. For most sub-K cooler architectures, the heat lift required at 2 K is a factor of $\sim$3 lower than that at 4 K, which is about the factor by which the 2-K lift is reduced relative to that at 4 K for a given compressor power consumed.

### 5.2 Telescope Design

The detailed design of the CALISTO telescope is a key aspect of our proposed study. An example configuration is our point design described in Goldsmith et al., 2008 [26], and shown in Figure 4. This design features a 4$\times$6-meter monolithic primary mirror used off-axis, and a secondary mirror which is deployed with a single hinge mechanism. This provides an optimal collecting area in a non-deployed primary mirror which fits into a 5-m fairing. While other materials could be considered, the baseline approach is to use silicon carbide (SiC), which is attractive given its favorable thermomechanical properties, and given it's success in the Herschel observatory, a system with comparable size and surface accuracy requirements to CALISTO. Other aspects are less clear, and there are several inter-related design choices that we propose to trade in our study, including:

1. On axis vs off-axis. As noted in Goldsmith 2008 [26], a benefit of the off-axis geometry is cleaner beams. However the off-axis construction will drive cost (delta to be studied), and some of the beam effects might be mitigated by insuring that all supports are cold and absorbing. The scientific impact of the two options should be carefully quantified.
2. Active vs passive. Given the progress in silicon carbide active mirror technology, and the cost and complexity associated with verifying the large-scale figure accuracy of a large cryogenic telescope, the lowest-cost, lowest-risk option may be a telescope which includes some on-orbit figure adjustment authority, either in the primary itself or in a smaller image of the telescope.
3. Cost vs telescope aperture. A key aspect of our submission to NASA and the 2020 Decadal survey should be the run of cost with telescope and system size.

## 6 Detectors and Instrumentation

To address the scientific goals oulined above, the primary instrumentation for CALISTO is a suite of 5–8 moderate-resolution (R$\sim$500) wideband spectrometers, which combine to span the full 35 to 600 $\mu$m range instantaneously with no tuning. The detailed arrangement of the modules in the focal plane and the degree to which multiple modules can couple to the same sky position simultaneously is a subject for the detailed study, but any given frequency channel will couple at least tens and up to 200 spatial pixels on the sky. These spectrometer approaches are described below, after an overview of the detector technology and system requirements for the readout.

Broadband imagers (cameras) are also possible on CALISTO, and this could be particularly powerful for the short wavelengths where the beam is small and the confusion limit is thus deep (Section 3). This will be addressed in the study, but since it does not strongly drive the detector performance or format, it is not discussed in this paper.

Higher-resolution spectroscopic capability is another topic that is under consideration, will be addressed in detail in the study, but is not discussed in this document. Possibilities include etalons or Fourier-transform modules which could be brought in front of the grating backends, both of which potentially offer an order of magnitude enhancement in spectral resolution. The former can be relatively compact but introduces a penalty for scanning. The latter preserves the fundamental sensitivity to within a factor of 2, but will be large, particularly at the long wavelengths.

Finally, we note the potential for heterodyne spectrometer arrays. While not benefitting from the cryogenic aperture, phase-preserving spectrometers offer the only means of obtaining velocity information and detailed line profiles for Galactic ISM studies as well as protostars and protoplanetary disks. As a guide to the sensitivity, the magenta curve increasing with frequency in Figure 2 shows the sensitivity of a quantum-limited receiver to at 10-km/s wide line. If the line profile itself is not of interest, and line confusion and line-to-continuum concerns are not a concern, then the direct detection system is more sensitive even at very narrow linewidths. These aspects will be addressed in the study.

## 6.1 Superconducting Micro-Resonator-Based Detector Arrays

Arguably the most important recent development for CALISTO is the progress in superconducting detectors based on high-Q resonators which can be multiplexed in the RF or microwave at high density ($\sim 10^3$ detectors per octave of readout bandwidth on a single line). This greatly reduces the complexity of the cold wiring, and with careful design, enables an observatory with a total far-IR pixel count in the hundreds of thousands to a million. For comparison, Herschel had a total of 3,686 far-IR direct detectors, and SPICA will have a comparable number (though at much greater sensitivity). In particular, the kinetic inductance detector (KID) relies on thin-film microresonators which change resonant frequency as quasiparticles created by absorbed photons shift the resonators' inductance [18, 41]. The frequency shift may be monitored by recording the complex (amplitude and phase) transmission of an RF or microwave tone tuned to the resonant frequency. The response is linear provided changes in the loading are small. Due to the high quality factors (narrow linewidths) that can be achieved, thousands of KIDs may be read out on a single RF/microwave feed line, using no cryogenic electronics except a single cold (e.g. ~20 K) microwave amplifier.

KID performance has steadily improved, and device sensitivities are now approaching the those of the SQUID-multiplexed bolometer systems in multiple groups worldwide (e.g. MUSIC [56] and NIKA / AMKID[49, 72]). The best reported sensitivities to date are $4 \times 10^{-19}\,\mathrm{W\,Hz^{-1/2}}$, more than sufficient for any ground-based or sub-orbital application. Further development is required to meet the requirements for CALISTO spectroscopy, but there are clear pathways to improving sensitivity for low backgrounds, namely by boosting the response with smaller-volume inductors, and increasing the effective quasiparticle lifetime through the use of suspended structures. The system-level aspects are also maturing, with scientific measurements now underway with KIDs at multiple telescopes. As an example, the MAKO project shown in Figure 5 is a 350 $\mu$m KID camera built by members of the Caltech / JPL detector group [64, 47]. It consists of 432-pixels read out with a single RF line, and is now operating very close to the the photon noise limit at the Caltech Submillimeter Observatory (CSO) (Figure 5).

While the KID uses the photo-response of the resonator's inductance, another approach is to use the its capacitance to measure the density of photo-produced quasiparticles via their tunneling rate from a reservoir in which the photons are absorbed. This is the basis of the quantum capacitance detector (QCD), with roots in the technology of quantum computing [57, 11, 62, 20]. The QCD is a naturally small-volume device that is already demonstrating optical NEPs down to $2 \times 10^{-20}\mathrm{W\,Hz^{-1/2}}$, meeting CALISTO's spectroscopy requirement, also shown in Figure 5.

## 6.2 Readout

With resonator Qs of $10^5$, 2000 devices can be arrayed per octave of readout bandwidth with negligible cross talk or frequency collisions. Assuming that a single RF line can carry 2 octaves (e.g. 100 MHz to 400 MHz), then this single line can service 4000 detectors. For each readout line, the KID or QCD readout consists of monitoring resonator frequencies with relatively straightforward if computationally-intensive signal processing algorithms. The most important question for CALISTO is the power consumption that will be required. The signal which interacts with the array must be digitized at ~500 Msamples per second, then Fourier transformed (FFT) at approximately the desired detector sampling rate, on order 1 kHz, so each FFT has on order 1 million points. The present Caltech implementation uses an FPGA on a ROACH[2] platform, no effort has yet been made to reduce power consumption for this ground-based pathfinders.

The path for CALISTO and other flight systems using this type of readout will be to develop a dedicated application specific integrated circuit (ASIC) which combines the digitization, FFT, and tone extraction in a single chip. Scaling from 7-bit ASICs that have been developed, the estimated power consumption for a 2-GHz, 12-bit system that would service the 2-octave band described above is conservatively ~1W. Thus we anticipate that on order half a million pixels could be read out for a couple hundred watts of total power, well within the budget of a large mission such as CALISTO. This is an important topic for development / demonstration in the coming years.

[2]Reconfigurable Open-Architecture Computing Hardware

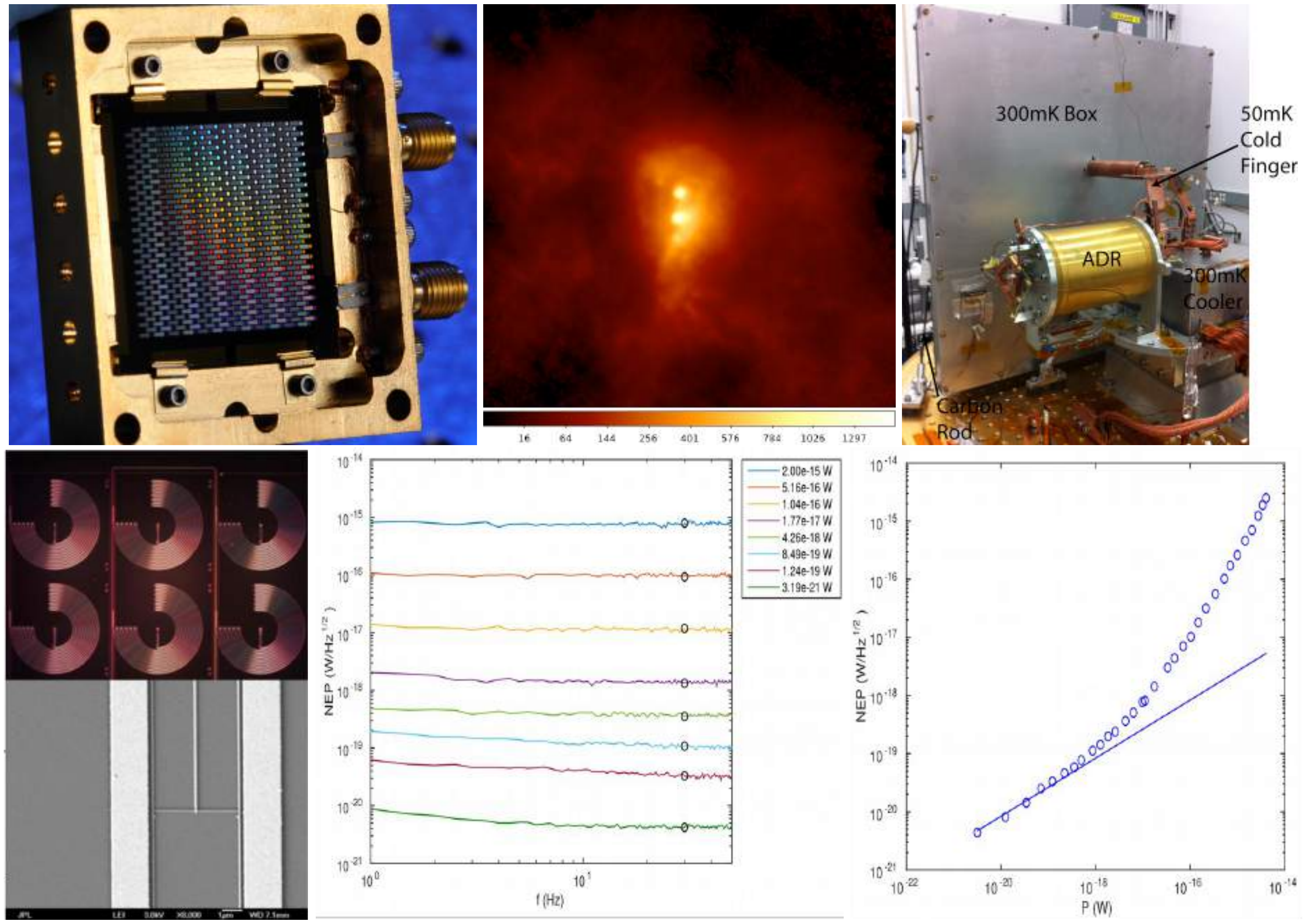

Figure 5: Sub-Kelvin Resonator-based detector technology. Above left shows the 432-pixel kinetic inductance detector (KID) array that forms the heart of the 350 $\mu$m MAKO camera, and (center top) an image of SGR B2 obtained with MAKO at the Caltech Submillimeter Observatory (CSO). Below shows views and measured noise performance of a quantum capacitance detector (QCD). Both devices can be multiplexed in groups of $\sim 10^3$ per readout line and thus are viable detector technologies for the few-$10^5$ total pixel counts we envision for CALISTO. The KID technology has demonstrated a high level of system maturity with the readout, optical coupling, and operation on sky, while the QCD is already showing photon-shot-noise limited sensitivity at the very low backgrounds required for CALISTO. Top right shows a prototype 300 mK / 50 mK cooling system which cools 10 kg with a flight-like mechanical suspension.

Finally, the system requires cryogenic low-noise amplification on each readout line. The Caltech laboratory system uses silicon-germanium transistor amps; they are currently operated at 4 K, but offer suitable noise temperatures at 20 K as well, so 20 K operation is feasible. As with the warm readout, little effort has been made to reduce power consumption of these devices; but amplifiers with good noise performance have been demonstrated with 700$\mu$W dissipation. A promising approach is a staged amplification which is integrated with the observatory cryogenic system: a low-power, moderate-gain stage at 20 K, combined with one or two higher-power, higher-gain stages closer to the warm side. At 1 mW per readout chain, the 125 amplifiers required for a 500 kpixel system would dissipate 125 mW, a tractable load for 20 K.

## 6.3 Spectrometer Modules

**Grating Spectrometers** For the short wavelengths ($\lambda < 200\,\mu$m), conventional first-order echelle gratings are a good choice, and each spectrometer will cover a bandwidth of 1:1.5, coupling to a planar 2-D array with ~200 spectral $\times$ 200 spatial pixels. Grating module sizes will range up to 30-40 cm, for example for a 130–200 $\mu$m module, with a mass less than 5 kg. An example grating module design is shown in Figure 6.

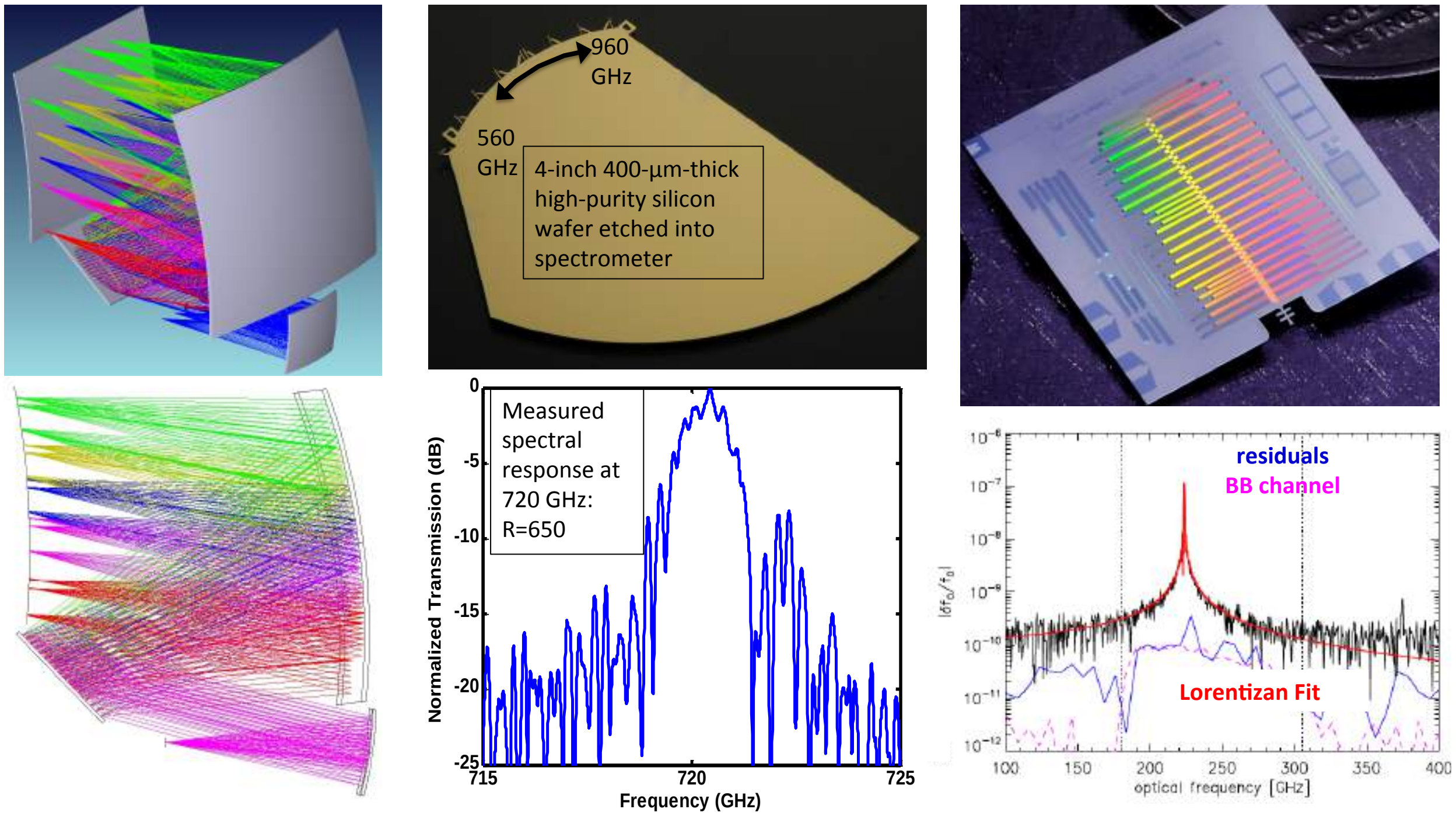

Figure 6: CALISTO spectrometer approaches. Left shows a conventional wide-band slit-fed echelle grating module as is envisioned for the short wavelengths. It processes a full 165-beam-long slit and a bandwidth of 1:1.5 at R=400 in a package which is $\sim$1800$\lambda$ on a side. At longer wavelengths, a more compact architecture is required, and has spurred development of two new approaches, both of which have demonstrated basic functionality. At center is a silicon-immersed waveguide grating spectrometer; its size is on order $\lambda \times R/3.4$. At right is a superconducting filterbank spectrometer (SuperSpec), which can be used at the lowest CALISTO frequencies. The prototype pictured has 80 spectral channels with R ranging from 200 to 800 and is 1 cm in size.,

**Silicon-Immersed Waveguide Spectrometers** For $\lambda > 200\,\mu\mathrm{m}$, conventional spectrometers become too large and bulky, so we will use waveguide spectrometers formed from high-purity float-zone silicon wafers. These devices have a size on order the resolving power $\mathrm{R} \times \lambda/\mathrm{n}$, where $n = 3.4$, the index of refraction of silicon. These spectrometers build on our success with Z-Spec[8], and we have demonstrated R=700 operation in such a device, demonstrating that the dielectric loss is not a concern. Each spectrometer couples a single beam, but since each is 2-dimensional, they can be stacked, with detectors then arranged in 2-D sub-arrays, each coupling a frequency sub-band for all of the spectrometers in the stack. As a example, a stack of 100 grating module for 230 to 360 $\mu$m could be achieved in a package $\sim$10 cm by 10 cm by 30 cm, with a mass of $\sim$6 kg or less.

**Superconducting On-Chip Spectrometers** For the longest-wavelength CALISTO bands, a superconducting chip-based spectrometer can be used. This technology consists of a filterbank circuit formed from superconducting transmission line lithographically patterned onto silicon with an integrated detector array. Because it is a path-folding device, the dimensions can be quite small, as the photograph in Figure 6 shows. A complete a 200-channel wideband spectrometer 'pixel' could be packed into a thin silicon die with surface area of few square centimeters, so the chips could be arrayed into a 2-dimensional focal plane with as many as a few hundred units. Development of these filterbank spectrometers is proceeding rapidly (see the SPIE papers [58, 2, 30, 59, 31]) and a ground-based demonstration is anticipated in the next 2 years. At present the devices use niobium as the superconductor, which limits the operation to $\lambda < 380\,\mu\mathrm{m}$, but higher frequency operation is possible with higher-temperature superconductors, for example NbTiN, which could extend down to 200 $\mu$m. A similar capability can be provided by the $\mu$-Spec system developed at Goddard [12], though this has a size similar to the silicon waveguide spectrometers for a given $\lambda \times R$ product.

## 6.4 Cooling of the Instrument

To enable the very low detector NEP, and insure that there is negligible optical loading from the instrument, the full spectrometers modules will likely be cooled to below 100 mK. No fundamental obstacles exist, as sub-100-mK cooling in space has been demonstrated in both Astro-H and Planck. The Astro-H soft X-ray calorimeter uses a multi-stage adiabatic demagnetization refrigerator (ADR) backed by a 1 K liquid helium bath in conjunction with closed-cycle 4-K class coolers [24, 60]. Planck used an open-cycle dilution refrigerator [66] in which both $^3$He and $^4$He are expended. However, with an estimated total sub-K mass approaching 100 kg, the system for CALISTO will be much larger than either of these previous implementations. While some aspects could be scaled, the use of consumables is likely to be prohibitive for the CALISTO system, and is undesirable as it limits the lifetime. A better approach will be a system similar to SPICA, in which the sub-K system is designed to interface with the facility 4K and 2K coolers described in Section 5.1.

On aspect that is immediately clear is that staging from the 2-K observatory heat sink, an intercept will be required at an intermediate temperature, e.g. 0.5 K. Multiple architectures are possible including closed-cycle dilution refrigerators [15], multi-stage adiabatic demagnetization refrigerators (ADR) [60], and hybrid coolers using $^3$He sorption and ADR, as is baselined for SPICA / SAFARI [14]. We refer the reader to a paper comparing these options (Holmes et al., 2010 [35]), but scaling from our laboratory demonstrations and calculations for the BLISS study [9], we estimate that the cooler elements could require $\sim$30% of the mass of the cold instruments, and that per 10 kg of cooled mass, they would require heat lifts at 5 mW at 4 K and 2 W at 1.7 K.

### 6.5 Data Rate

Ideally, the CALISTO system would be able to store fully-sampled data from all detectors at unit duty cycle. Assuming 16 bits at 100 Hz for 250 kilo-pixels creates a total raw rate approaching 0.5 Gbit per second, or 35 TBits per day. This is larger than currently-planned L2 missions which use Ka band DSN (e.g. Euclid plans 0.85 Tbits / day). Thus some form of on-board compression should be considered. Unlike optical / near-IR missions which point and stare, in the far-IR the approach is to scan map or modulate at some frequency, so on-board processing will will require new algorithms, for cosmic-ray removal and map-making / demodulation.

Optical communications are a promising solution to the CALISTO downlink challenge. The higher gain provided by the shorter-wavelength translates into a large increase in data rate for a given mass and power relative to a Ka band system, and this technology has been progressing steadily. In the last 2 years, NASAs Lunar Laser Communication Demonstration (LLCD) demonstrated successful laser communications including downlink at 622 Mbits / sec between a satellite in lunar orbit, the Lunar Atmosphere and Dust Environment Explorer (LADEE), and ground stations on the Earth [5]. Optical communications is being pushed by the Planetary Division, and is featured in the coming call for Discovery mission proposals. L2 is particularly well-suited to optical communications, since L2 is always in the night sky. A baeline design, consistent with optical-communications development targets begins with an existing concept for a Deep-space optical Transceiver (DOT) that is now baselined for the Discovery mission – it is essentially a 22-cm telescope coupled to a few-W laser. The transmit power required depend on the collecting area of the receiver. NASA is considering a 12-meter class receiver on the timescale of 2025 to support of deep space communications, but this would probably not be required for CALISTO at L2. For 1 Gbit / sec at L2, a 1-meter receiver requires 14 W of transmit power, but with a 3-meter receiver, the transmit power is a more reasonable 1.6 W (William Farr, personal communication), making the full system less on order 100 W including the actuation. Thus dedicated 3-meter class receivers at 1–2 sites could achieve data rates in excess of 14 Tbits/day with only 4 hours of downlink, corresponding to the full data rate from CALISTO with only modest on-board compression.

## 7 Cost Landscape

CALISTO was studied by JPL Team-X in various exercises between 2005 to 2008. The telescope configuration described above, the associated cryocoolers, the deployed sunshade, an allocation for instruments, and operations for a 5-years mission were estimated to cost $1.7 billion (FY2008$). The breakdown is provided in Table 4. Of course, re-assessing this is an important aspect of our proposed pre-decadal study. One key point is that we are now advocating substantially more capable instrumentation for CALISTO. While the new frequency-domain multiplexing schemes naturally enable the large formats, we nevertheless expect that the increased scope will increase both the instrument and science terms in the budget (over the full mission life) relative to the 2008 estimate.

Team-X also considered lower-cost options, ranging from reducing the aperture to a (Herschel-like) 3.5-meter circular telescope to eliminating some of the instrumentation (reducing cryogenic mass and data rate). For the lowest-cost of these, which combined both reductions, the estimate cost was $1.1 B ($FY06). Finally we note for comparison the as-built costs for the Herschel ($1.1 Billion, per ESA), and Planck ($700 M) missions.

Table 4: CALISTO JPL 2008 Cost Estimate Breakdown

| Item | Cost [$M ‘08] |
|---|---|
| Management, Systems Eng., Mission Assurance | 101 |
| Payload System (primarily science instruments) | 196 |
| Flight System (incl. sunshield, telescope, coolers) | 608 |
| Operations and Ground Data System | 132 |
| Launch Vehicle | 156 |
| Assembly, Test and Launch Operations | 53 |
| Science | 114 |
| Education, Public Outreach | 6 |
| Mission Design | 10 |
| Reserves | 330 |
| Total Estimated Project Cost | 1,706 |

# The Dusty Co-evolution of Black Holes and Galaxies:
# A Science Case for a Large Far-IR Space Telescope

A whitepaper written in response to the COPAG call for large astrophysics missions to be studied by NASA prior to the 2020 Decadal Survey

L. Armus (lee@ipac.caltech.edu)
Infrared Processing and Analysis Center, Caltech

P.N. Appleton (IPAC), C.M. Bradford (JPL), T. Diaz-Santos (UDP), C.C. Hayward (Caltech), G. Helou (IPAC), P.F. Hopkins (Caltech), M.A. Malkan (UCLA), E.J. Murphy (IPAC), A. Pope (UMASS), B. Schulz (IPAC), H. Teplitz (IPAC)

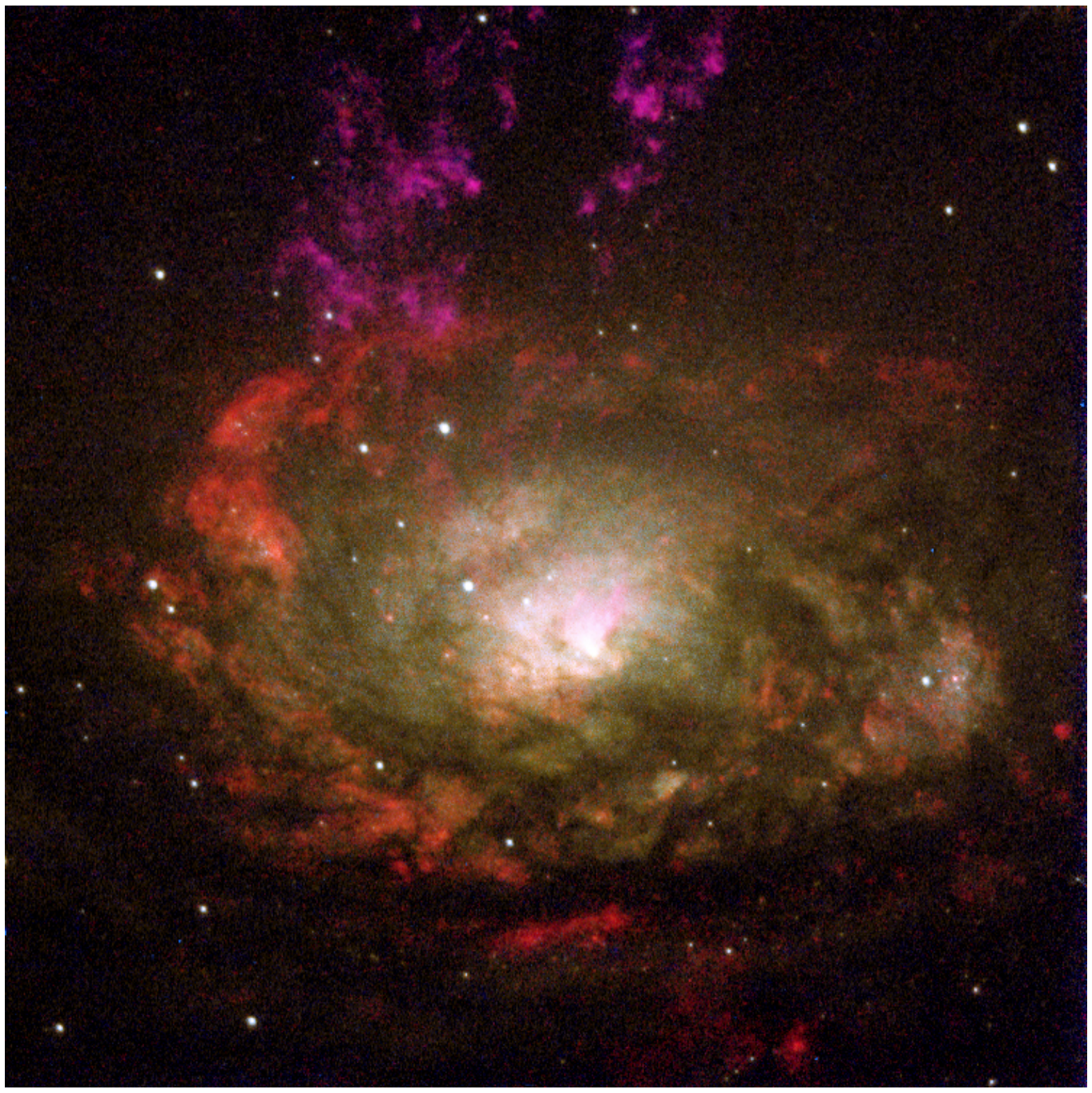

***Cover Image**: Hubble Space Telescope of the nearby Circinus galaxy. The dusty center shows evidence for a massive black hole, a powerful starburst, and outflows of hot gas.*

## Background & Key Questions

In order to obtain a comprehensive picture of galaxy evolution, we need to accurately measure the growing population of stars and super-massive black holes in galactic dark matter halos. This evolution is determined by a complex interplay of physical processes (gravity, gas heating and cooling, star formation, black hole fueling, and feedback from star formation and AGN) that couple on scales ranging from < 1pc to tens of Mpc.

One of the most striking results to appear in the last decade has been the discovery that the mass of the central black hole and the stellar bulge in galaxies are correlated [11,17]. The idea that galaxies spend most of their lives on a star formation vs. stellar mass "main sequence" [48] further suggests that star formation and black hole accretion are intimately linked. Understanding how this relationship is built over time drives a great deal of observational and theoretical astrophysics, providing considerable motivation for the next generation of ground and space-based observatories. Despite the success of cosmological simulations that model the hierarchical growth of galaxies [7, 34, 35], and observations suggesting that periods of significant AGN accretion occur during episodes of enhanced nuclear star-formation [6, 9, 23], a number of critical questions still remain, such as: **When do the first heavy elements appear, and how does the chemical history of the Universe regulate the collapse of the first stars and the build-up of galaxies? How and when do the first black holes form and how does the black hole – bulge mass relation evolve with redshift for galaxies on and off the star forming main sequence? How and when does feedback from stellar winds, supernovae and AGN regulate star formation and the growth of galaxies?**

Although we have broadly measured the evolution of the bolometric luminosity density to z~3, the relative contribution of AGN and star formation at early epochs is quite uncertain. To piece together a complete picture of the co-evolution of galaxies and black holes requires the ability to make extremely sensitive infrared measurements of the most obscured regions at the centers of faint, distant galaxies.

## The Need for Background-Limited FIR Spectroscopy

More than half of all the light emitted from stars is absorbed by dust and re-emitted in the infrared [8]. While traditional UV and optical diagnostics can be severely hampered by dust attenuation, FIR spectroscopy provides a direct measure of the basic physical properties (density, temperature, pressure, kinematics) of the ionized ($T\sim10^4$ K), the neutral atomic, and the warm (T~100-500 K) molecular gas in obscured galaxies. **It is the only part of the electromagnetic spectrum that gives a complete picture of all phases of the interstellar medium, from atoms to complex organic molecules.** The infrared is rich in fine-structure lines of Oxygen, Carbon, Nitrogen, Neon, Sulfur and Silicon covering a wide range in ionization potential, as well as molecular hydrogen and dust (Polycyclic Aromatic Hydrocarbons - PAHs). Together, these features constrain the strength and hardness of the interstellar radiation field [18, 3, 20]. This is extremely relevant since z~3, UV-selected galaxies seem to have starbursts with harder radiation fields, higher ionization potentials and/or different abundances than those at z~0 [50]. The FIR lines can be used to trace molecular outflows [37, 38] and infer the size of the starburst [39], and mid-J transitions of CO can distinguish starburst from AGN heating of the molecular gas [4, 30, 40, 41]. A FIR spectroscopic survey of high-redshift galaxies can solidly establish the history of early chemical enrichment, the rise of metals, and the presence of organic molecules.

With ISO, Spitzer and Herschel we have studied large samples of dusty galaxies in the local Universe [5, 2, 9, 42, 45, 46], identified PAHs in the most luminous galaxies out to z~4 [24, 19, 43] and detected the populations responsible for the bulk of the FIR background at z~1 [16]. However, our knowledge of how AGN and galaxies grow together, and the role of feedback in rapidly evolving, dusty galaxies at z >2-3 is extremely limited.

**In order to produce a complete census of AGN and chart the growth of super-massive black holes and stellar mass in dusty galaxies across a significant fraction of the age of the Universe, a broadband, FIR spectrometer capable of reaching the natural astrophysical background over the ~30-300μm range is required.** FIR cooling lines in z~2 IR galaxies should have fluxes ~$10^{-19}$ $Wm^{-2}$. The rest-frame MIR lines will be 5-10x fainter. JWST will provide our first glimpse of the earliest galaxies, yet most of the mid-infrared diagnostic lines will pass out of the observable range of the JWST spectrographs by z~2. ALMA is already detecting z >5-6 galaxies [31, 44, 47, 49], yet it operates in limited atmospheric windows, and cannot access the rest-frame MIR spectral features. In particular, we require: **(1)** sensitivity of ~$1x10^{-20}$ $Wm^{-2}$ in an hour to detect normal dusty galaxies at z > 2 and luminous galaxies at z > 4, **(2)** broad spectral coverage from ~30-300 μm to cover the key redshifted MIR and FIR lines, **(3)** a spectral resolving power of R > 100 to separate individual atomic features from dust emission and absorption, and **(4)** spectral multiplexing to place 10-100 beams on the sky and allow for significant samples to be built up rapidly. The required sensitivity and wavelength coverage is impossible to reach from the ground, but could be achieved with a large, actively cooled telescope in space.

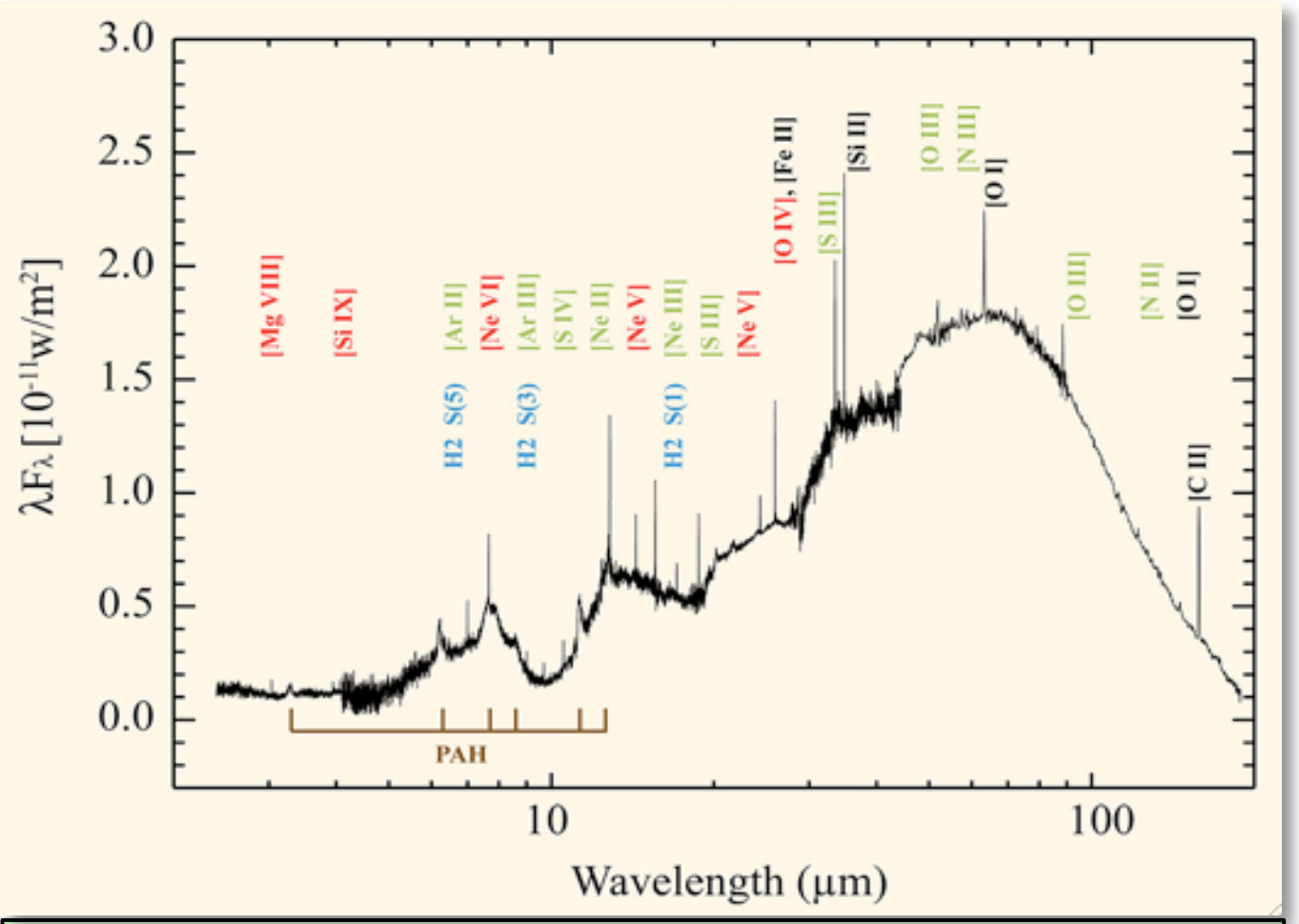

Infrared spectrum of the Circinus galaxy showing a strong dust continuum studded with features from small grains, ions, and warm molecular gas [21].

CALISTO, a cold T~4K, 5m class telescope which has been put forward for the FIR Surveyor concept (see Bradford et al. whitepaper), is the only mission currently envisioned for the next decade capable of achieving the goals outlined above. Through FIR spectra of thousands of distant galaxies, CALISTO will allow us to map out the history of galactic chemical enrichment, accurately estimate the bolometric fraction contributed by AGN and starbursts in even the most obscured sources, and trace AGN and stellar feedback via IR absorption and emission features providing a complete census of the buildup of galaxies and black holes over the past 10 Gyr.

## Far-Infrared Spectral Line Studies of the Epoch of Reionization

Asantha Cooray (UC Irvine; accooray@uci.edu), James Aguirre, Phil Appleton, Matt Bradford, Caitlin Casey, Phil Mauskopf, Bade Uzgil

Existing cosmological observations show that the reionization history of the universe at $z > 6$ is both complex and inhomogeneous. All-sky CMB polarization measurements provide the integrated optical depth to reionization. A detailed measurement of the reionization history may come over the next decade with 21-cm HI radio interferometers, provided that they are able to remove the foregrounds down to a sub-hundredth percent level. Deep sky surveys, especially those that employ gravitational lensing as a tool, are now efficient at finding Lyman drop-out galaxies at $z > 7$, though with a large systematic uncertainty on the redshift due to degeneracies between Lyman drop-out and dusty galaxy SED templates. During the next decade the study of reionization will likely move from studying galaxies during reionization from $z = 6$ to 9 to understanding primeval stars and galaxies at $z > 10$.

In the JWST era current studies that focus on the rest-frame UV and optical lines to study the ISM and gas-phase metallicities of galaxies at $z \sim 1$ to 2 will quickly extend to $z$ of 6. In the post-JWST era a far-infrared space telescope with a factor of 10 sensitivity improvement over SPICA will enable studies on the gas properties, AGN activity and star-formation within galaxies at $z = 6$ to 15, in addition to a large list of sciences at $z < 6$. This redshift range is especially important for our understanding of the cosmic origins, formation of first stars, galaxies and blackholes, and the onset of large-scale structure we see today. The community is already struggling with many scientific issues during this epoch. For example, an outstanding problem involves the growth of supermassive blackholes and the presence of billion to ten billion solar masses backholes at $z > 6$ at an age of 600-800 Myr after the Big Bang. One possibility to grow such high masses is seed blackholes associated with massive PopIII stars. Could we directly observe the formation of such massive stars at $z \sim 12$ to 15? And could we study the blackhole activity in galaxies at $z \sim 7$ to 10 as these blackholes grow in mass rapidly to values measured at $z \sim 6$?

To aid study of reionization 20 to 600 $\mu$m spectroscopic observations can: (a) disentangle the complex conditions in the ISM of $z = 6$ to 15 galaxies by measuring the gas densities and excitation, and the prevalence of shock heating; (b) compare the conditions of the ISM in high-redshift galaxies with local galaxies to address whether faint dwarf galaxies found at low-redshifts are analogues of galaxies during reionization; (c) use spectral line diagnostics to study AGN or star-formation regulated actively within first galaxies, including the formation of first massive blackholes; and (d) detect, measure, and map out molecular hydrogen rotational line emission from primordial cooling halos that are the formation sites of first stars and galaxies at $z > 10$.

The role of far-infrared spectral capabilities will allow diagnostic studies and ways to establish the role of feedback, radiation, and AGNs, among many others, in regulating star-formation in reionization era galaxies. For example, [OIV]26 and Ne[V]14.3 are high-ionization lines that are enhanced in AGN environments and far-IR diagnostics such as [OIV]26/[SIII]33 or [NeV]14.3/[NeIII]15.6 ratios provide a direct measure of the AGN fraction to galaxy luminosity, even when there is signicant dust extinction. Such diagnostics then allow a way to distinguish galaxies at $z > 8$ that harbor AGNs and are likely to grow to optically luminous quasars detected with wide sky surveys such as SDSS. The mid-IR to far-IR spectral region provides both the depth and the range to initiate a wide array of studies that are still limited to lack of developments in the observational capabilities.

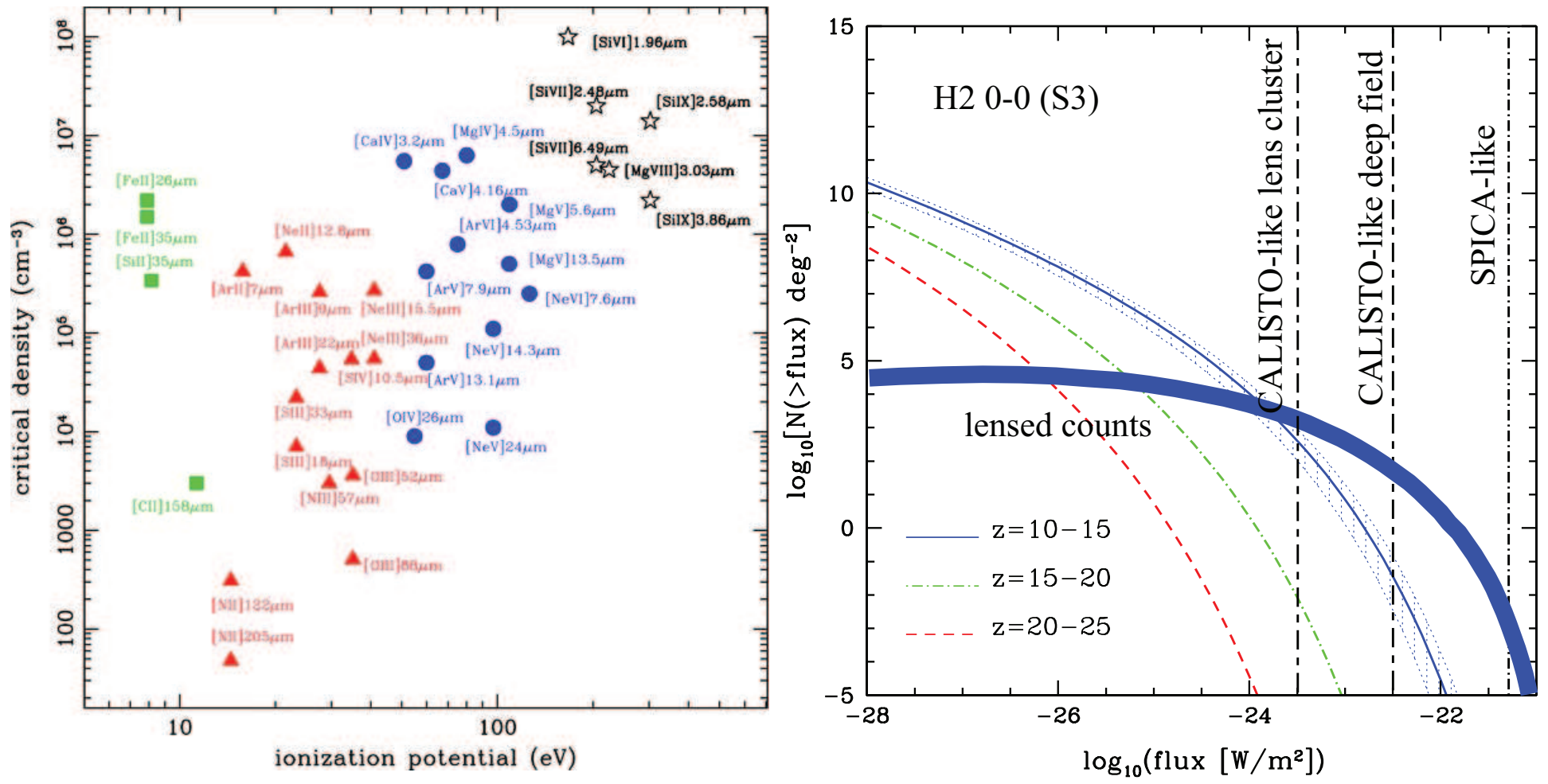

Figure 1: *Left panel:* Density-ionization diagram (Spinoglio et al. 2009) for far-IR spectra lines. Color coding shows lines separated to different conditions and radiation fields, such as stellar/HII regions or AGNs. *Right panel:* $H_2$ 0-0 S(3) cumulative number counts as a function of the flux density (Gong et al. 2012). In addition to intrinsic counts, we also show the gravitationally lensed counts at $z > 10$ by foreground galaxies. While SPICA or SPICA-like mission will not have the sensitivity to detect molecular hydrogen in mini-halos at the onset and during reionization, a far-infrared mission with at least a factor of 10 sensitivity improvement over SPICA, such as CALISTO, will be necessary to detect many of the important molecular hydrogen lines in the rest-frame mid-infrared wavelengths. This will require a deep 1 deg$^2$ survey over 2000 hrs. Another factor if ten depth can be achieved, on average, lensing cluster survey, similar to Hubble Frontiers Fields.

Moving to the highest redshifts molecular hydrogen is now understood to be the main coolant of primordial gas leading to the formation of very first stars and galaxies. It is also the most abundant molecule in the universe. There is no other signal from primordial gas cooling at the earliest epochs in either the low-frequency 21-cm background or any other cosmological probe that the community has considered. **Molecular hydrogen cooling in primordial dark matter halos will then likely the only tracer to study the transition from dark ages at $z > 20$, when no luminous sources exist, to reionization at $z < 10$.** At a metallicity $Z \sim 10^{-3.5}$ $Z_\odot$ gas cooling will transit from $H_2$ to fine-structure lines. At $z < 8$, when primordial molecular hydrogen is easily destroyed by UV radiation, the prevalence of shocks in the ISM may provide ways to form a second and later generations of molecular hydrogen. The rotational lines of molecular hydrogen span across a decade of wavelength from 2 to 20 $\mu$m. JWST will study molecular hydrogen out to $z$ of 1 and SPICA may be able to study them to $z$ ∼3 to 4, but at $z > 6$ SPICA does not have the required sensitivity (Fig 1 right panel).

Even if not individually detected, a far-IR survey telescope will use intensity fluctuations, similar to power spectra in CMB and Cosmic Infrared Background but in 3D due to spectral line redshift mapping, to study the spatial distribution of $H_2$. This intensity mapping technique also relies on a cross-correlation with a second line of $H_2$ from the same redshift interval to minimize foreground line contamination. The requirements for $z \sim 6$ far-IR fine-structure and $z > 10$ $H_2$ mapping of primeval cooling halos are 20 to 600 $\mu$m spectral coverage and a noise level below $10^{-22}$ W/m$^2$ in a deep 1000 to 2000 hour integration over a sq. degree area. CALISTO is one step in this direction.

# Mapping Turbulent Energy Dissipation through Shocked Molecular Hydrogen in the Universe

A whitepaper written in response to the COPAG call for large astrophysics missions to be studied by NASA prior to the 2020 Decadal Survey

P. N. Appleton (apple@ipac.caltech.edu)
NASA Herschel Science Center, Caltech
(626-395-3119)

L. Armus (IPAC), C. M. Bradford (JPL), G. Helou (IPAC), P. Ogle (IPAC), A. Cooray (UC Irvine), J. Aguirre (U.Penn), Caitlin Casey (UC Irvine), Phil Mauskopf (ASU), Bade Uzgil (U. Penn/JPL), P. Guillard (IAP, Paris), F. Boulanger (IAS, Orsay)

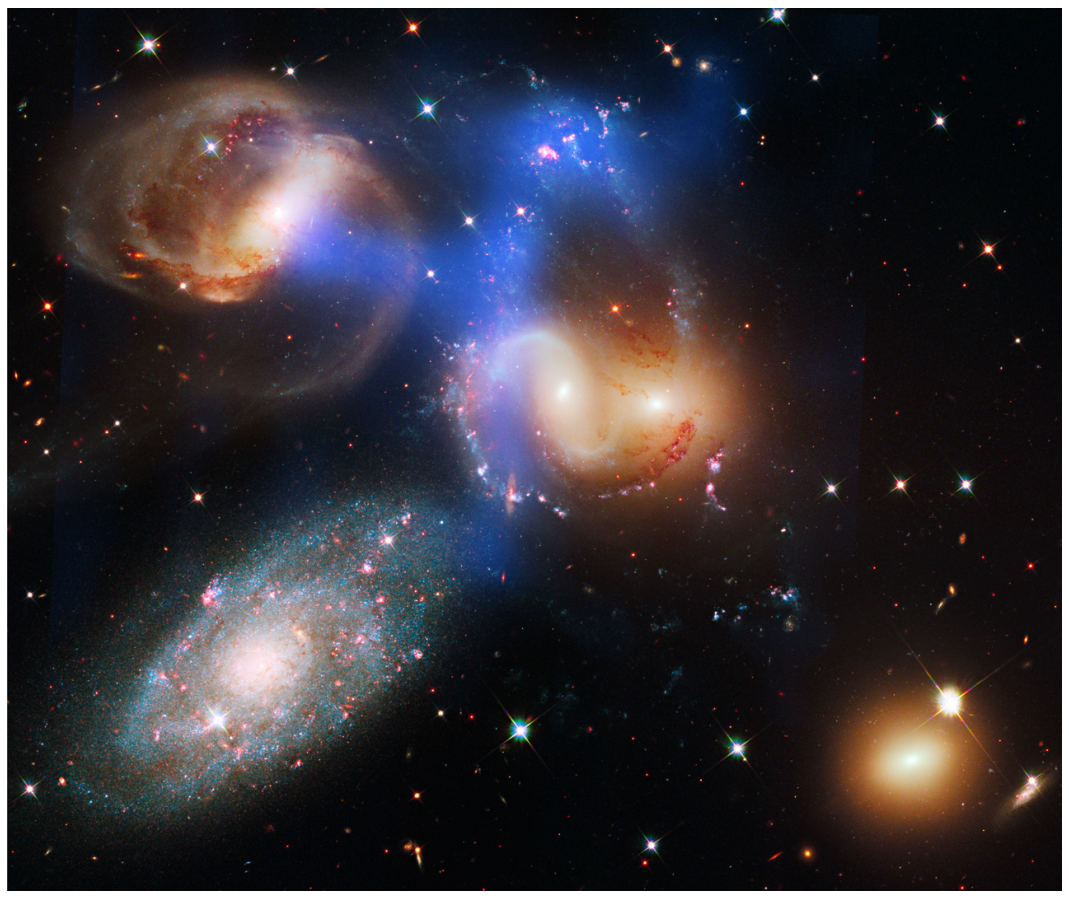

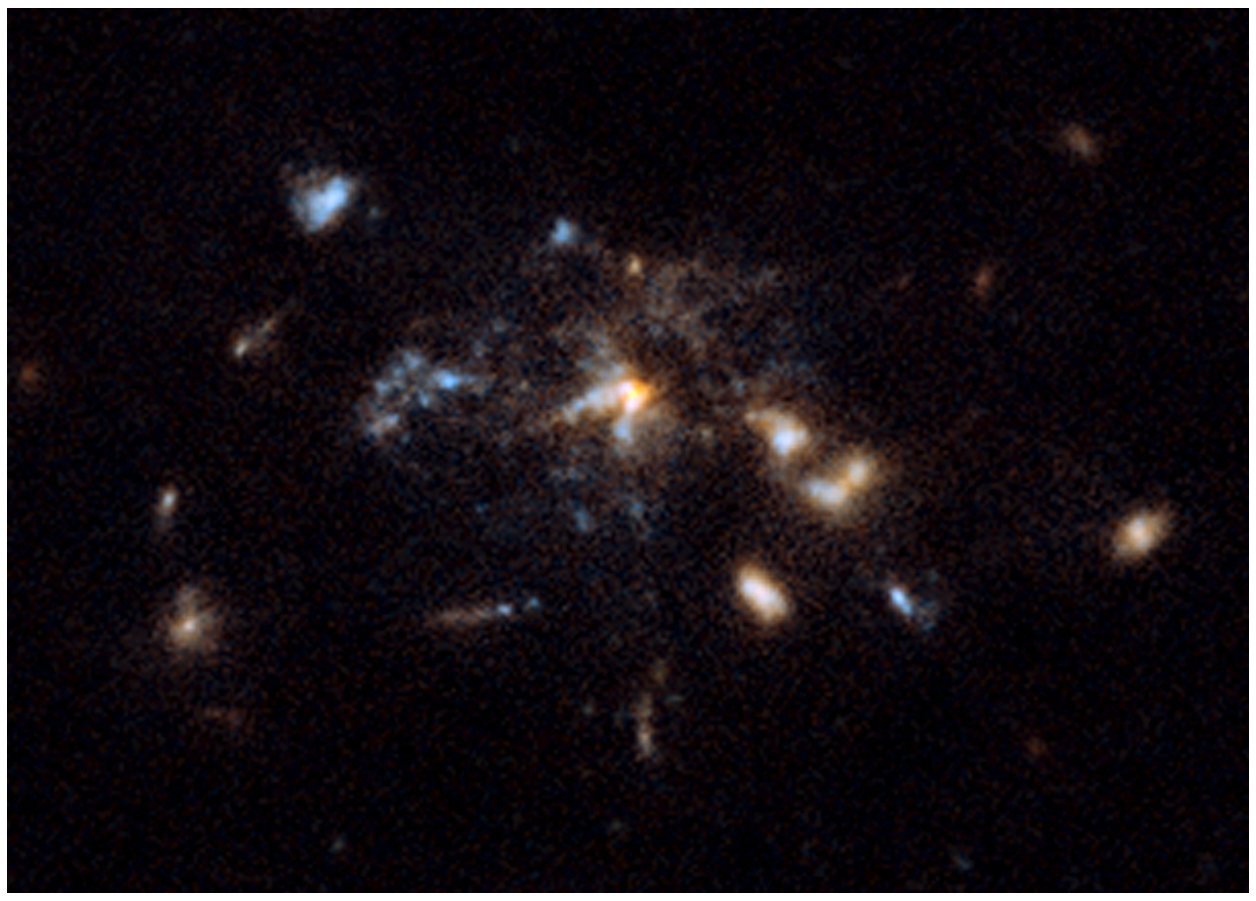

**Cover Images:** ***(Left)*** *Shock-excited 0-0S(1) molecular hydrogen (blue) emission from Stephan's Quintet,* ***(Right)*** *The most extreme warm $H_2$ emitter found by Spitzer just before it ran out of cryogen--The "Spiderweb" proto-cluster at $z = 2.16$.*

**Background and Motivation:** Probing the growth of structure in the Universe is arguably one of the most important, yet uncharted areas of cosmology, ripe for exploration in the next few decades. Molecular hydrogen ($H_2$ and HD), along with the first heavy metals born in the first supernovae, played a vital role in cooling the primordial gas (e. g. Santoro & Shull 2006), setting the scene for the formation of first large-scale baryonic structures. The IGM enrichment by heavy elements also led to the formation of dust, which in turn almost certainly led to a rapid acceleration of $H_2$ formation on grains for redshifts z < 15 (Cazaux & Spaans 2004). Almost all the primary cooling channels for gas at z > 2 occur in the far-infrared/sub-mm bands, including dust and Polycyclic Aromatic Hydrocarbons (PAH) emission, the mid-IR rotational lines of molecular hydrogen (e. g. 0-0 S(3)9.7μm, S(1)17μm, S(0)28μm), and the far-IR lines of [O I]63μm, [Si II]34.8μm, [Fe II]25.9μm and [C II]157μm. **The far-IR is therefore a critical window for the study of the initial growth and evolution of gas in the universe over cosmic time.**

During the *Spitzer* mission, it was discovered that there exists a population of galaxies exhibiting extremely strong emission from warm (typically 100 < T < 500 K) molecular hydrogen (Ogle et al. 2010). One of the most striking examples was found in the giant intergalactic filament in Stephan's Quintet (Appleton et al. 2006, Cluver et al. 2010), where the mid-IR molecular hydrogen lines were unusually bright (Cover page). This warm molecular gas is believed to be tracing the dissipation of mechanical energy in shocks (Guillard et al. 2009) and turbulence, caused by the collision of a high-speed intruder galaxy with a tidal filament. $H_2$ emission dominates the gas cooling in the Quintet's filament, being enhanced relative to other important coolants (Appleton et al. 2013). Thus molecular hydrogen seems to be a powerful coolant, even in the local universe where metals are more abundant than in the early universe. Other nearby examples have also been found, where the $H_2$ appears to be heated by collisions between galaxies (Peterson et al. 2012, Cluver et al. 2013, Steirwalt et al. 2014). Furthermore, Ogle et al. (2010) showed that 20% of nearby 3CR radio galaxies also showed excessively high warm $H_2$ emission, most likely from shocks caused by the passage of the radio jets through the host galaxy (see also Nesvadba et al. 2010; Nesvadba et al. 2011). Guillard et al. (2012) demonstrated that radio galaxies exhibiting strong HI outflows also showed similar characteristics. In some cases, the warm molecular hydrogen provides clues about the suppression and removal of gas in the inner regions of galaxies containing AGN (Ogle et al. 2014). **Studying emission from warm molecular hydrogen can provide a direct measure of the properties of the gas cooling, which sets limits of timescale for the dissipation of turbulent energy. This is likely to be important for understanding the physical conditions that lead to negative ISM feedback on star formation in the universe.**

**Bridging to the high-redshift Universe:** Before *Spitzer* ran out of cryogen, it detected a number of very powerful $H_2$-emitting galaxies, including several central cluster galaxies (e. g. Zw 3146 at z = 0.3; Egami et al. (2006)), where the $H_2$ line-luminosity is an order of magnitude brighter than those seen in individual galaxy collisions. Shocks and or cosmic ray heating (Guillard et al. 2015; Ferland et al. 2008) may be responsible for some of these large luminosities, but by far the most powerful warm $H_2$ emitting system was detected by Ogle et al. (2012) in the z = 2.15 radio galaxy proto-cluster PKS1138-26 (knows as the "Spiderweb": cover page). The luminosity in a single $H_2$ rotational line (the 0-0 S(3) 9.66μm), was a phenomenal $3 \times 10^{10}$ $L_{\odot}$, 100 x brighter than Stephan's Quintet. The existence of such extreme $H_2$ emitters begs the question of whether $H_2$ could be used to probe turbulence in the early universe (see Appleton et al. 2009). The molecular hydrogen lines therefore represent an important window into turbulence that can only be explored in the far-IR. Although JWST's mid-IR capability will allow the study of the nearby universe in the higher-excitation $H_2$ lines, **the exploration of $H_2$ in the low-lying rotational lines (which traces the dominant mass and cooler temperatures) will impossible beyond z > 2, without a large cool FIR telescope in space.**

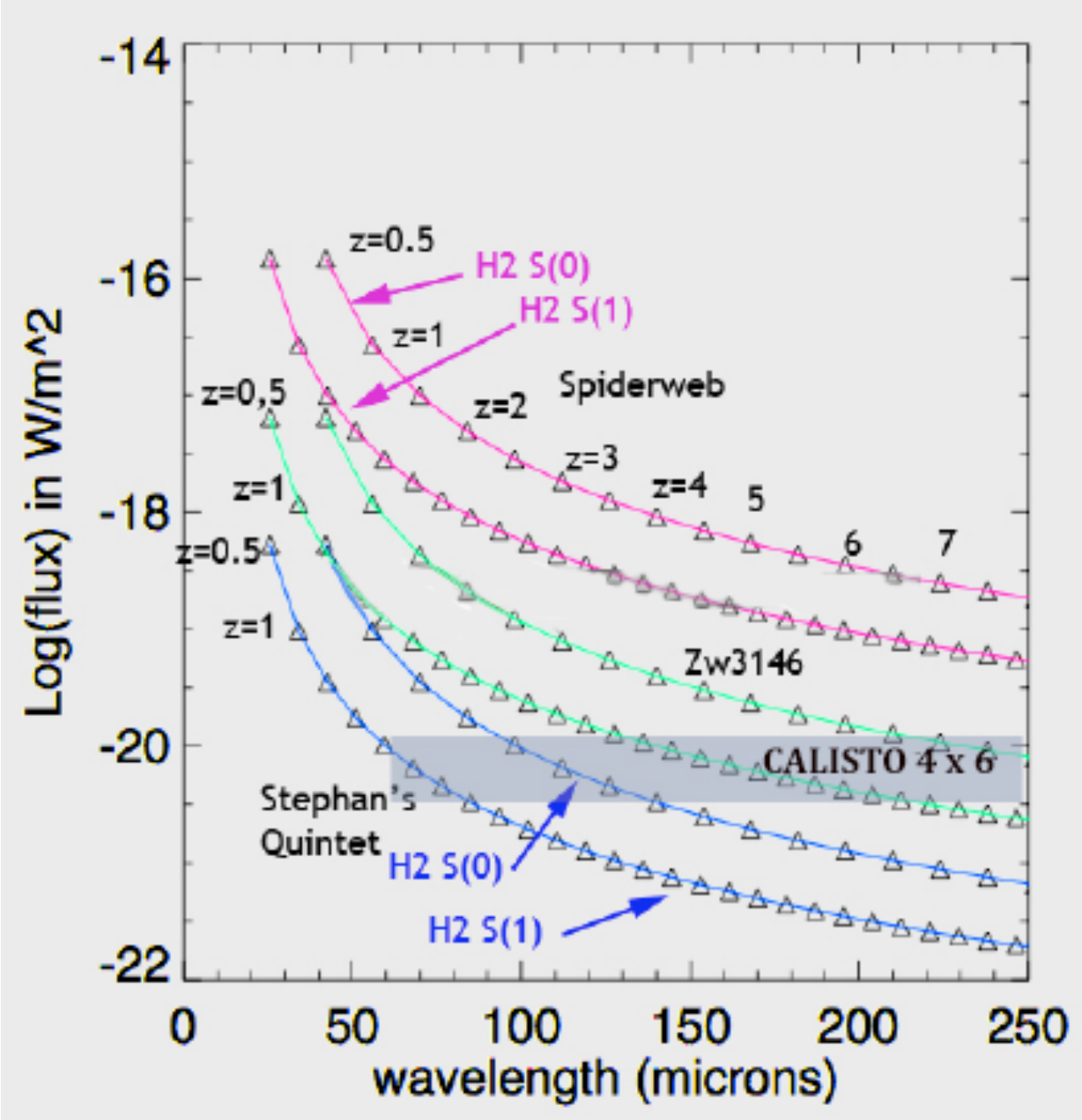

*Estimates of the 0-0S(0)28μm and 0-0S(1)17μm ground-state pure-rotational $H_2$-line fluxes ($W\ m^{-2}$) for the Spiderweb (PKS1138-26) and the central cluster galaxy in Zw 3146 shifted in increments of z= 0.5 as a function of observed wavelength.*

*The grey box shows the achievable sensitivity of the CALISTO telescope with the 4 x 6 element spectrometer discussed by C. M. Bradford in an associated white paper. These sources, if they exist at higher-z, would be readily detected at z > 5-6. Compact group sources like Stephan's Quintet could be studied to z > 1.*

Although the detection of individual proto-galaxies at redshifts > 10 are probably beyond the reach of current instrumentation (see Appleton et al. 2009), the detection of powerful clusters at z > 4 is quite feasible (see figure). These systems will provide an important insight into energy dissipation and galaxy formation in the most over-dense regions in the universe. CALISTO, a cold T~4K, 5m class telescope which has been put forward for the FIR surveyor concept (see Bradford et al. whitepaper), is the only mission currently envisioned for the next decade capable of detecting the low-excitation $H_2$ gas that we associate with large-scale turbulence. Extra sensitivity could be gained by mapping around strong lensing systems, to dig deeper, and to avoid foreground confusion. This would allow exploration of limited volumes of the high-z universe to greater depth. Potentially ALMA-Band 10 has a capability of reaching few x $10^{-20}$ W $m^{-2}$ in long integrations. However, the tiny primary beam (5 arcsecs at 850 GHz), and narrow fractional band-width (< 0.3%) would make the detection of shocked-enhanced primordial gas extremely difficult, requiring *a priori* knowledge of the precise target location and redshift. CALISTO, on the other hand, can potential detect turbulent $H_2$ out to high redshift in many $H_2$ lines simultaneously because of its huge wavelength grasp. In addition, its larger beam would allow efficient mapping, especially if more than one beam is placed on the sky simultaneously (the 4 x 6 concept of Bradford et al.). At the highest z, the best way to detect primordial gas may be through the method of intensity mapping (e. g. Gong et al. 2013), where a CALISTO-like spectrometer could be used to map spatially and exploit spectrally, the faint statistical signals of proto-galaxies at z > 10. **A cold FIR telescope in space would provide a vital probe of heating and cooling processes at work in the youngest galaxies, greatly expanding NASA's portfolio, and providing a unique suite of tools for studying the Cosmic Dawn.**

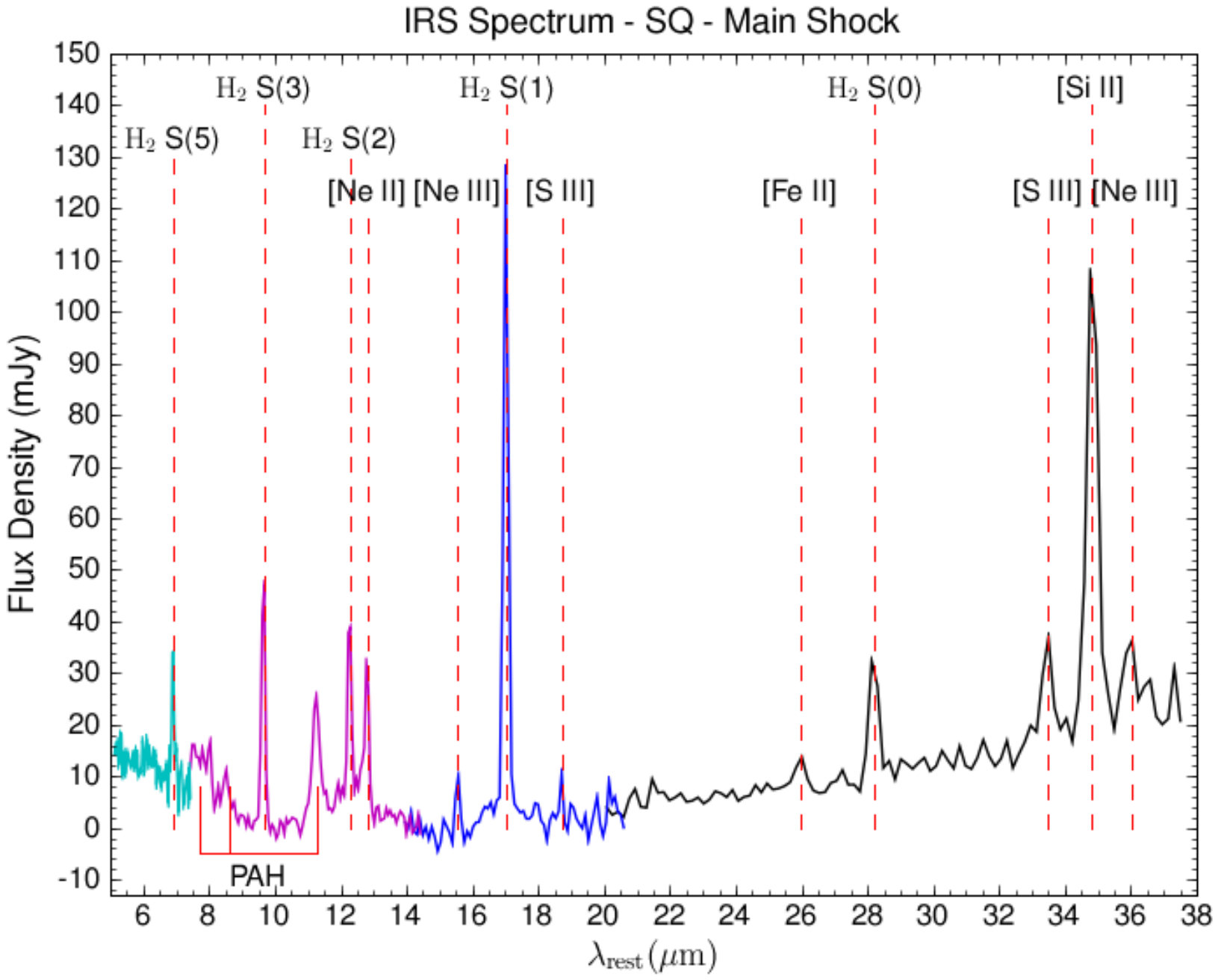

The IRS spectrum of the turbulent shock structure in the Stephan's Quintet Compact Group (Appleton et al. 2006; Cluver et al. 2010). The warm $H_2$ gas dominates the power from the region. [CII]157μm (Appleton et al. 2013) and [SiII] emission are the next most powerful line coolants. These lines are redshifted into the far-IR and sub-mm at high-z.

# – Dust in Distant Galaxies –
# Overcoming Confusion Noise with a 5m FIR Facility

Caitlin Casey, Matt Bradford, Asantha Cooray, James Aguirre, Phil Appleton, Phil Mauskopf, Bade Uzgil

The vast majority of galaxy evolution studies in the past fifty years have focused on deep optical and near-infrared ($\lambda < 5\mu m$) datasets, tracing galaxies direct emission from starlight. Yet half of all energy emanating from these galaxies is emitted in the far-infrared and submillimeter, where dust and gas emit[1,2]. Dust absorbs emission from young, hot stars and re-radiates that energy at long wavelengths, peaking at rest-frame ≈100$\mu$m. This dust emission provides very important clues to galaxies' evolutionary history, but is virtually unconstrained observationally over the majority of the Universe's evolution and in normal, Milky Way type galaxies.

Previous limitations in far-infrared instrumentation and the atmospheric opacity at these wavelengths has made detailed studies of dust in distant galaxies extremely challenging in the past, with only a handful of missions successfully surveying the sky in the past thirty years. These include the *IRAS*[3] (1983) and *ISO*[4] (1995) missions, and in more recent history *Spitzer*[5] (2003), *AKARI*[6] (2006), and *Herschel*[7,8] (2009). However, these missions were all significantly limited by a combination of limited sensitivity and small apertures, thus large beamsizes and confusion noise. While improving on detector sensitivity has been quite successful in the past few decades, overcoming confusion noise has been difficult.

Here we outline the impact that a 5m space-borne FIR facility would have on the direct detection of dust in distant galaxies at ≈50–200$\mu$m. With a relatively modest increase in aperture size, the confusion-limited depth is vastly increased over that of the *Herschel Space Observatory.* This is a simple consequence of the fact that at these frequencies, the 5-meter class aperture is reaching below the knee in the luminosity function, and the shallow faint-end slope translates to a rapid increase in depth with decreasing beam size.

---

Confusion noise arises when the density of sources on the sky is high relative to the beamsize of observations. Overcoming confusion noise is difficult without a large aperture. For example, the *Herschel* PACS and SPIRE instruments (operating at 70–160$\mu$m and 250–500$\mu$m, respectively) were confusion limited such that integrating for long periods of time would not improve the depth of the instruments surveys because the resolution was not sufficient to distinguish sources from one another. Strictly speaking, the confusion limit for a given facility, $\mathrm{S_{conf}}$, is the limiting flux density for which $\Omega_{\rm beam} \times N(> S_{\rm conf}) = 1$, where $\Omega_{\rm beam}$ is the solid angle of one beam (in deg$^{-2}$) and $N(> S_{\rm conf})$ is the density of sources at or above $S_{\rm conf}$ at the given wavelength. Confusion noise will dominate for sources with fluxes fainter than $S_{\rm conf}$, where there are more than one source per beam. Another commonly used qualification of confusion noise, used to derive confusion noise in existing observational datasets[9], defines $\mathrm{S_{conf}}$ as $\int_0^{x_c} x^2\,dn$, where $x$ is the measured flux, $x = S\,f(\theta,\phi)$, $S$ is the source flux convolved with the normalized beam response, $f(\theta,\phi)$, and $dn$ is the differential source distribution. In both cases, it is clear that the beamsize is the primary limitation in conducting very sensitive, deep FIR surveys.

Figure 1 illustrates the best measured differential number counts[10] at 70$\mu$m (from *Spitzer* MIPS[11,12] and *Herschel* PACS[13]), 100$\mu$m (from ISOPHOT[14,15,16] and *Herschel* PACS[13,17]) and 250$\mu$m (from BLAST[18,19] and *Herschel* SPIRE[20,21,22]). The differential number counts represent the number of sources per flux bin per area, plotted here in units of $dN/dS$ [mJy$^{-1}$ deg$^{-2}$], and is often fit to a parametric double power law or Schechter function, although it should be noted that such parametrizations are physically meaningless, as flux density is a function of luminosity, redshift and SED shape (dust emissivity, opacity, temperature, etc). Here we have overplot some best-fit double power law parametrizations, which extend to very low flux densities well below the limit of past FIR surveys. We have designated uncertainty on the faint end slope, $\alpha$, of the number counts to mirror the uncertainty in the data in that regime.

The right panels on Figure 1 show the cumulative number counts in units of sources per beam. For each panel, the left y-axis represents the beamsize of a proposed 5m FIR facility, while the right y-axis represents the beamsize of *Herschel*, a 3.5m facility. Solid horizontal lines illustrate the $S_{\rm conf}$ limit of one source per beam, as per the formal definition of confusion noise, while the dotted lines represent a more practical confusion limit of $^1/_4$ source per beam, in line with measured confusion limits from *Herschel* (note this value will depend strongly on the clustering of galaxies, which differs by wavelength). **What this shows us is that a beamsize that is reduced by a factor of $\sim$2 (due to the increased aperture of a 5m facility) will push the confusion limit at $\approx$70$\mu$m a factor of $\sim$3$\times$ deeper, and a factor of $\sim$10$\times$ deeper at 100$\mu$m and 250$\mu$m.** For example, the measured confusion limit at 100$\mu$m from *Herschel* PACS[17] is $\approx$0.15 mJy, which from Figure 1, appears to correspond to a cumulative number of sources per beam (right y-axis) of $\sim$0.13. Assuming the same effective limit with a 5 m facility (left y-axis value of 0.13), we derive a confusion limit at 100$\mu$m of $\sim$11$\mu$Jy. See Table 1 for our estimates at other wavelengths. The factor of ten improvement in the confusion limit at 100$\mu$m is due to the shallow slope of the faint-end of the number counts below 0.1mJy. While a steep slope would result in a less advantageous jump in the confusion limit, we know such slopes are unphysically possible as they would imply the cosmic infrared background (CIB[2]) should be several times larger than it is measured to be.

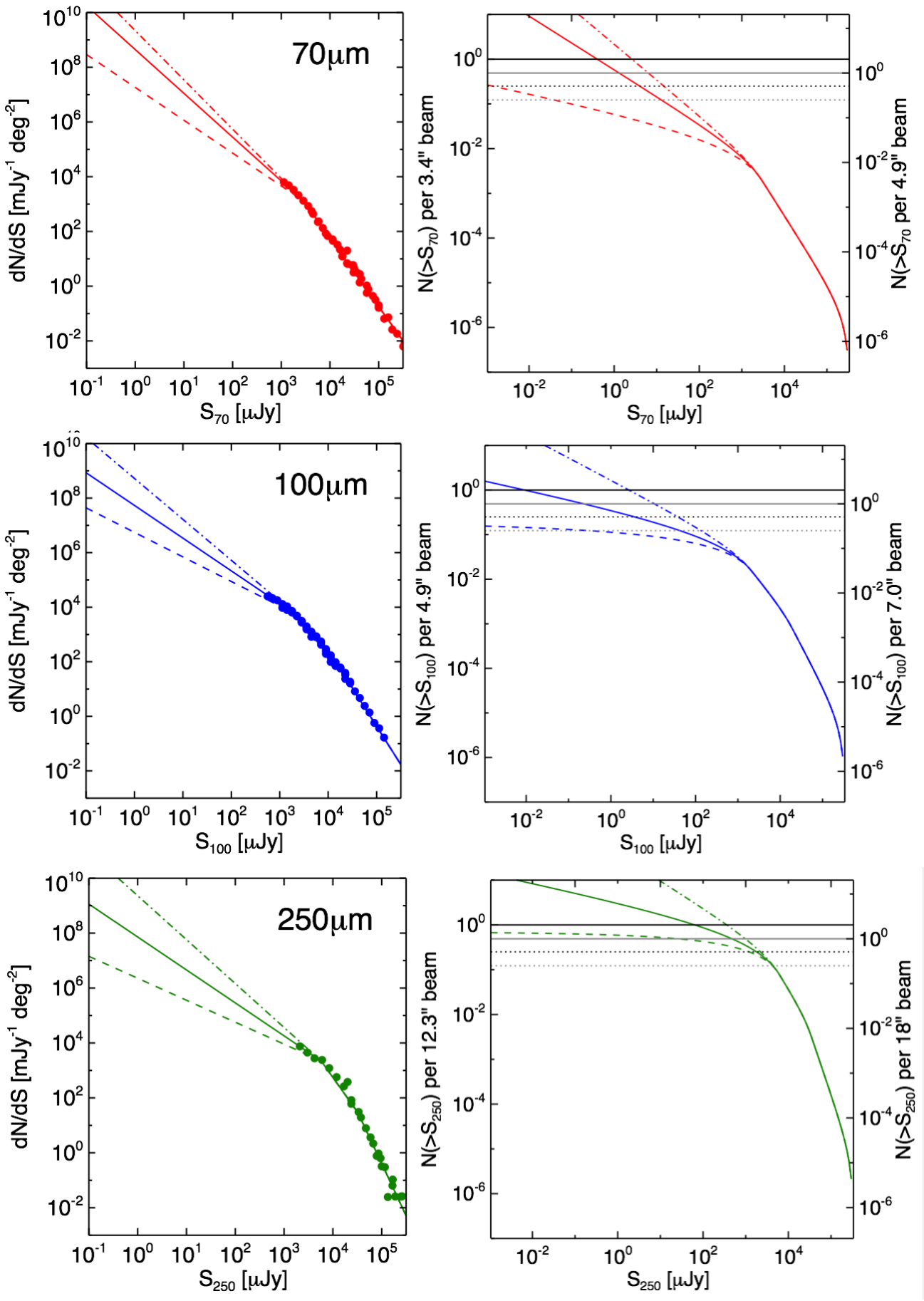

Figure 1: Differential and cumulative number counts at 70–250$\mu$m; see text for details.

So what is the scientific value of having a facility with such a low FIR confusion limit? A factor of ten in the confusion limit translates to a factor of ten improvement in the depth of FIR surveys, implying easy detection of Milky Way type galaxies in direct dust emission out to $z \sim 1.5$. The dramatic improvement in depth also implies the number of galaxies with direct detections in the FIR will increase by a factor of $\sim$100, extrapolating from the underlying shape of the dusty galaxy luminosity function[10]. This will allow very detailed analysis of dust emission, obscuration, and star-formation in distant galaxies on a far larger scale than has previously been possible and resolving the vast majority of individual galaxies contributing to the CIB, well below the knee of the galaxy luminosity function.

| Wavelength | *Herschel* conf. lim. | 5m conf. lim. | Factor of Improvement |
|---|---|---|---|
| 70$\mu$m | 35 $\mu$Jy | 11 $\mu$Jy | 3.2 |
| 100$\mu$m | 150 $\mu$Jy | 11 $\mu$Jy | 14 |
| 250$\mu$m | 460 $\mu$Jy | 68 $\mu$Jy | 7 |

**Table 1.** Estimated confusion limit for a 5m FIR facility in comparison to *Herschel*.

## Unlocking the Secrets of Planet Formation with Hydrogen Deuteride

Edwin A. Bergin (University of Michigan)

The following is a white paper that discusses how a far-IR observatory, with significant gains in sensitivity over Herschel, could make a major contribution towards our understanding of the physics of planet formation and the birth of habitable worlds.

*The Uncertain Gas Mass of Planet-Forming Disks*

Planets are born within disk systems (protoplanetary disks) that are predominantly molecular in composition with a population of small (sub-micron to cm sized) dust grains that represent the seeds of Earth-like planets. The most fundamental quantity that determines whether planets can form is the protoplanetary disk mass; forming planetary systems like our own requires a minimum disk mass of order $\sim$0.01 $M_\odot$ (i.e. the minimum mass solar nebula or MMSN; Weidenschilling, 1977; Hayashi, 1981). Estimates of disk masses are complicated by the fact that the molecular properties of dominant constituent, molecular hydrogen, lead it to be unemissive at temperatures of 10 – 30 K that characterizes much of the disk mass (Carmona et al., 2008).

To counter this difficulty astronomers adopt trace constituents as proxies to derive the $H_2$ mass. By far, the primary method is to use thermal continuum emission of the dust grains. At longer sub-mm/mm wavelengths the dust emission is optically thin probing the disk dust mass. With an assumed dust opacity coefficient, along with the ratio of the dust to gas mass, the disk gas mass is determined from the dust mass (Beckwith et al., 1990; Andrews & Williams, 2005). With this method the gas mass estimates range from $5 \times 10^{-4} - 0.1$ $M_\odot$ (Williams & Cieza, 2011). However, a variety of sensitive observations have demonstrated that grains have likely undergone growth to sizes 1 mm to 1 cm (at least) in many systems (Testi et al., 2014). Thus the dust opacity is uncertain and the gas-to-dust ratio is likely variable (Draine, 2006; Isella et al., 2010). The alternative is to use rotational CO lines as gas tracers, but these are optically thick, and therefore trace the disk surface temperature, as opposed to the midplane mass. The use of CO as a gas tracer then leads to large discrepancies between mass estimates for different models of TW Hydrae, the closest gas-rich disk (from $5 \times 10^{-4}$ $M_\odot$ to 0.06 $M_\odot$), even though each matches a similar set of observations (Thi et al., 2010; Gorti et al., 2011).

These uncertainties are well known with broad implications regarding the lifetime where gas is available to form giant planets, the primary mode of giant planet formation, either core accretion or gravitational instability in a massive disk (Hartmann, 2008), on the dynamical evolution of the seeds of terrestrial worlds (Kominami & Ida, 2004; Ida & Lin, 2004), and the resulting chemical composition of pre-planetary embroyos (Öberg et al., 2011). Given current uncertainties, we do not know whether our own solar system formed within a typical disk (Williams & Cieza, 2011). This extends beyond our planetary system as the frequency of extra-solar planet detections has been argued to require higher disk masses (Greaves & Rice, 2010; Mordasini et al., 2012).

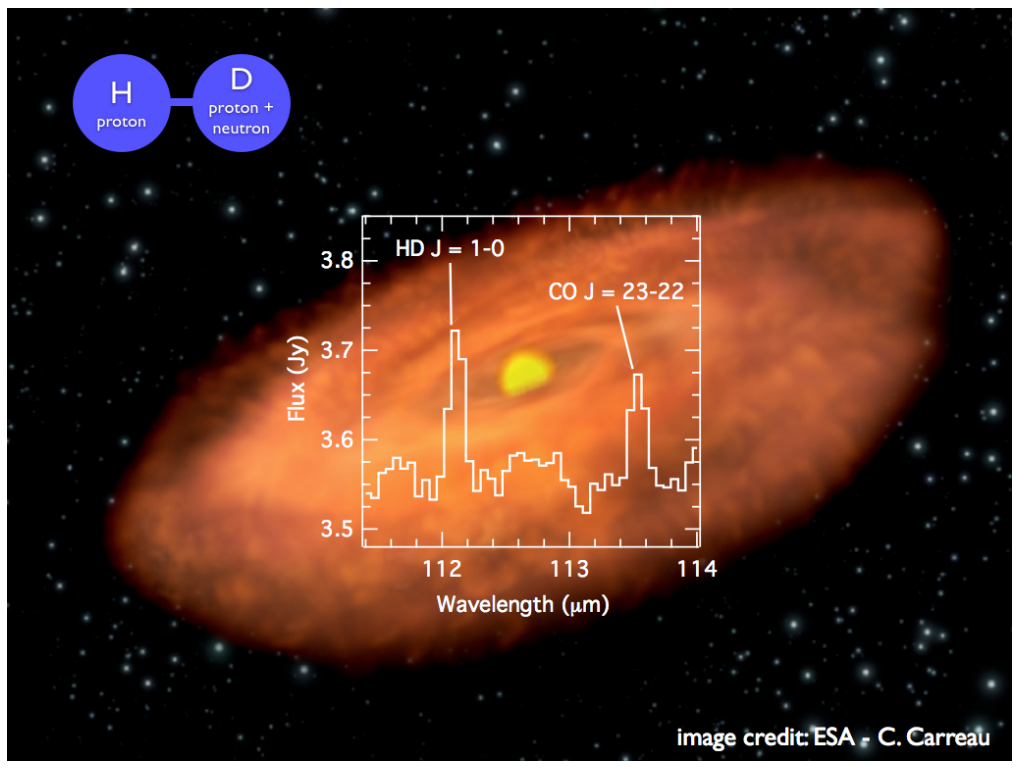

Figure 1: Herschel detection of Hydrogen Deuteride in the TW Hya protoplanetary disk superposed on an artist conception of a young gas-rich disk.

*Far-IR Spectroscopy, HD, and Disk Gas Masses*

Bergin et al. (2013), using the Herschel Space Observatory, detected the fundamental rotation transition of HD

at 112 $\mu$m emitting from the TW Hya disk (shown in Fig. 1). The atomic deuterium abundance relative to $H_2$ is well characterized to be $3.0 \pm 0.2 \times 10^{-5}$ in objects that reside within $\sim$ 100 pc of the Sun (Linsky, 1998), such as TW Hydra. Unlike carbon monoxide, HD and $H_2$ are only weakly bound on the cold (T $\sim$ 10 – 20 K) dust grains that reside in the mass carrying disk midplane (Tielens, 1983). Thus HD resides primarily in the gas throughout the disk with a known abundance relative to $H_2$. With energy spacings better matched to the gas temperature and a weak dipole the lowest rotational transition of HD is a million times more emissive than $H_2$ for a given gas mass at 20 K. It is therefore well calibrated for conversion of its emission to the $H_2$ gas mass in the disk offering the best chance to derive accurate disk gas masses in regions that are potentially actively forming planets. In the case of TW Hya the gas mass is estimated to be $> 0.05$ $M_\odot$, or many times the MMSN.

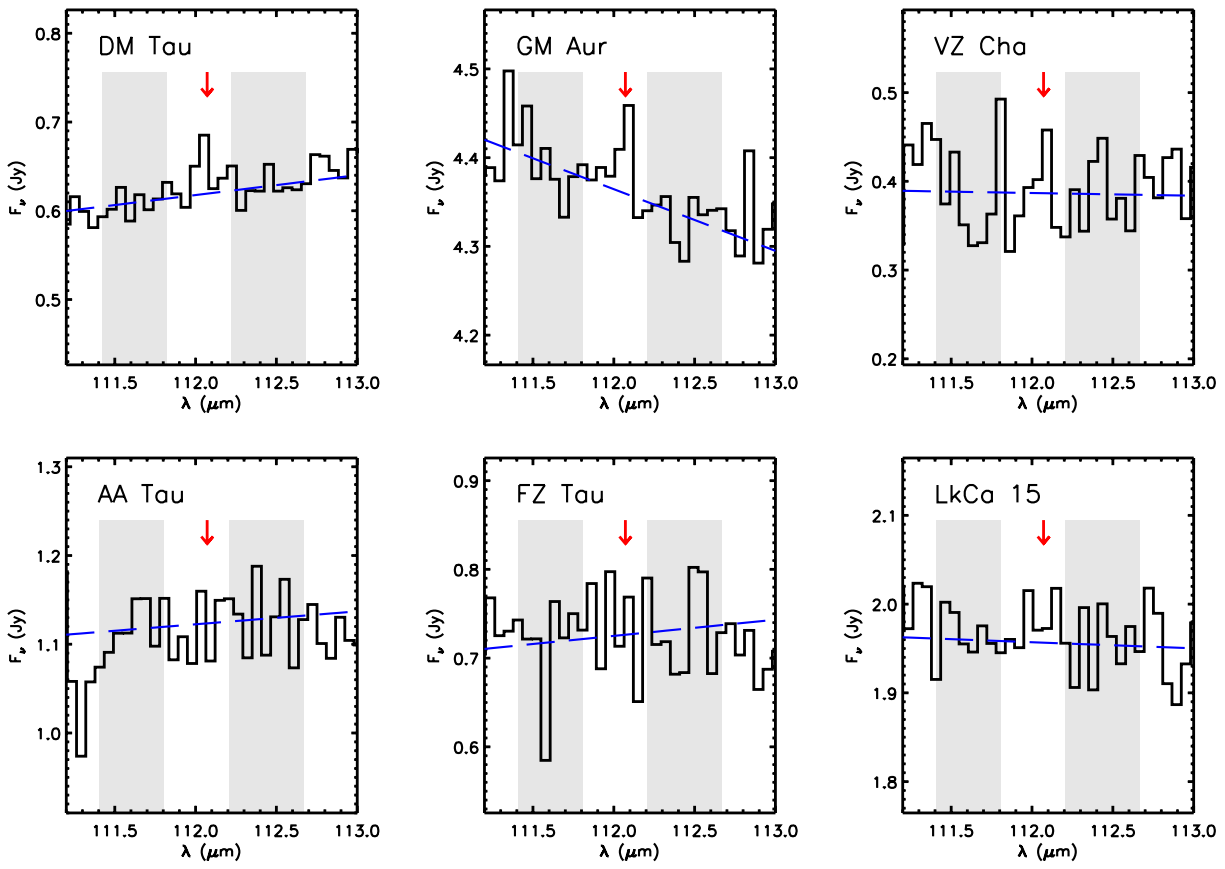

Figure 2: Results from the shallow survey of HD emission to be published by McClure et al. 2015, in prep.

Due to Herschel's limited lifetime the only other deep HD observations were obtained in the Cycle 1 program that resulted in the TW Hya detection. These observations, which are less sensitive than the TW Hya data, are shown in Fig. 2. For the most part, at this sensitivity limit, HD was not detected, although these disk are $\sim$3$\times$ more distant than TW Hya. However, marginal detections ($> 3\sigma$) were obtained in DM Tau and GM Aur hinting at the future promise for a high impact survey with a future far-IR facility.

*A survey of HD emission, can only be enabled with a sensitive Far-IR observatory. To move beyond the $\sim$ 3 systems with accurate gas masses, and open up our understanding of planet formation, we need to detections in* $>$ 100 *disk systems. This will provide the missing - and grounding - information on the gas masses of planet-forming disks.* Such a survey of a hundred of the nearest systems can determine the timescales of planet formation, whether $H_2$ is present in debris disk systems, and set needed constraints for disk dynamical models. A large telescope might also *resolve* HD in the closest systems, allowing for constraints to be placed on the uncertain gas density profile.

Knowledge of the disk mass also breaks the degeneracy between disk mass and chemical abundance. As an example, Favre et al. (2013) used HD with $C^{18}O$ finding that the CO abundance is more than an order of magnitude below that in the dense ISM. This was explored more directly (Du, Bergin, and Hogerheijde 2015, in prep.) using a complete thermochemical model (Du & Bergin, 2014) to analyze CO isotopologue data but also Spitzer/Herschel observations of water vapor. This work finds that to match observations, the abundance of elemental oxygen and carbon must be reduced in the upper layers by orders of magnitude. This missing carbon and oxygen must reside as ices in the dense midplane locked inside pebbles or even planetesimals. This information is crucial as the Atacama Large Millimeter Array is now providing resolved images of gas tracers, such as CO and other species. Without HD in TW Hya we would assume that readily accessible gas tracers (e.g. CO, HCN, etc) suggest that the gas mass is low, while instead it is the beginnings of planet formation that is being revealed. Thus there is tremendous synergy of a future far-IR facility with ground based instruments; only the far-IR can provide this fundamental information.

## A FIR-Survey of TNOs and Related Bodies

J. M. Bauer[1,2], P. F. Goldsmith[1], C. M. Bradford[1], A. J. Lovell[3]

[1]Jet Propulsion Laboratory, California Institute of Technology, Pasadena, CA, USA; [2] Infrared Processing and Analysis Center, California Institute of Technology, Pasadena, CA,USA; [3]Department of Physics and Astronomy, Agnes Scott College, Decatur, GA, USA

The small solar-system bodies that reside between 30 and 50 AU are referred to as the Trans Neptunian Objects, or TNOs. They comprise, in fact, the majority of small bodies within the solar system and are themselves a collection of dynamically variegated subpopulations, including Centaurs and Scatter-Disk Objects (SDOs), as well as "cold" (low-inclination and eccentricity) and "hot" (high eccentricity) classical Kuiper Belt populations (KBOs; Gladman et al. 2008). These minor planets are the reservoir of the comets that routinely visit our inner solar system, the short period comets, and so cloud the distinction between asteroids and comets. They are primordial material, unmodified by the evolution of the solar system and are the sources of volatile materials to the inner solar system (Barucci et al. 2008). Study of TNOs can thus inform us about the early history of the solar system, and how its composition has evolved over the time since it was formed.

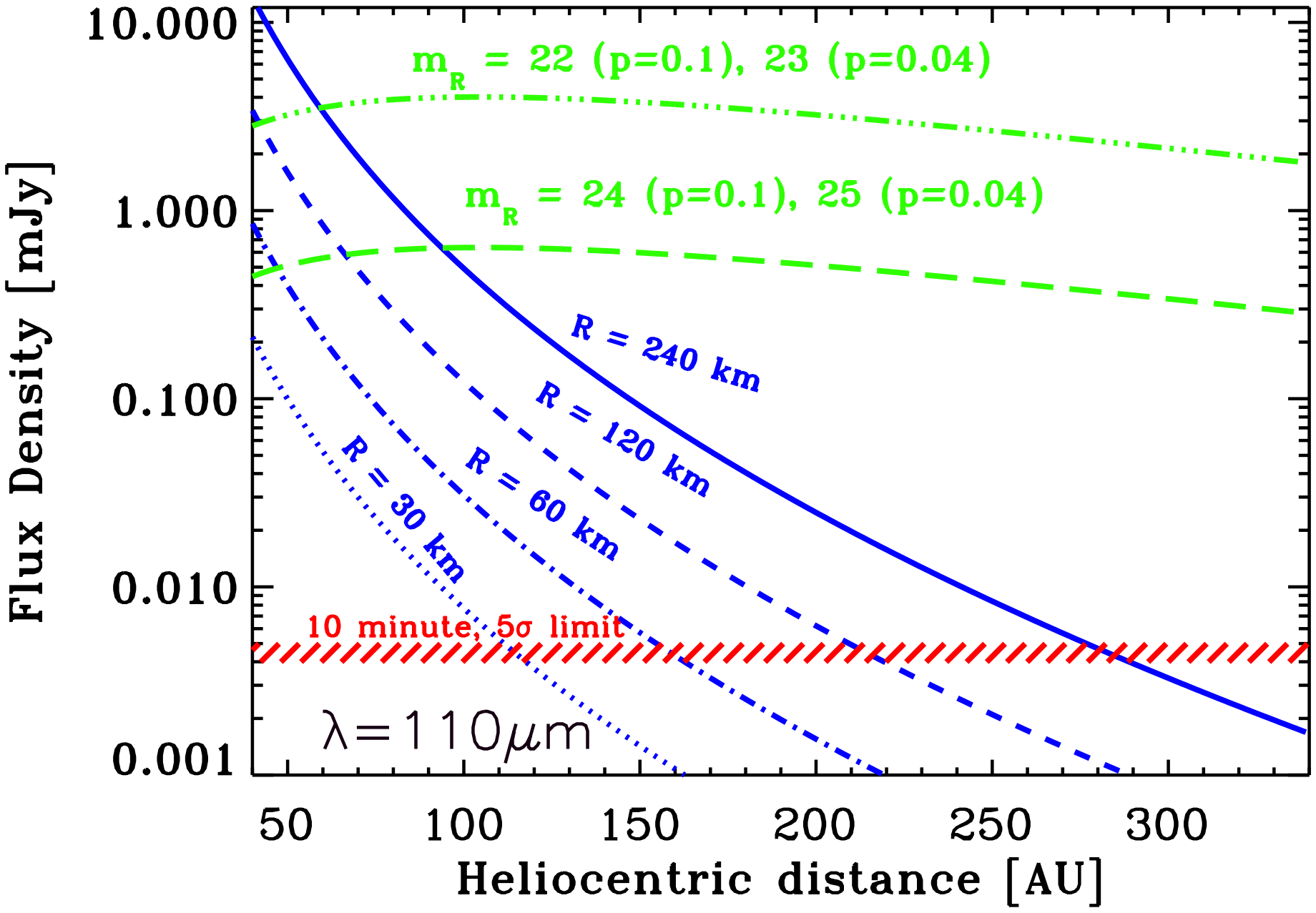

**Figure 1:** Flux density at 110 μm from TNOs of different radii (curves labeled by radius R) compared with CALISTO's 5σ detection limits for integration times of 10 min (diagonal-dashed horizontal line) The curves with indicated R-band (620 nm) magnitude $m_R$, and geometric albedo, p, give the flux from TNOs which are at limit of optical detectability. These lie well above the CALISTO limits.

*A FIR TNO Survey*: Surveys of these more distant solar system bodies to date are limited by optical-band sensitivities down to the 22-24 magnitude level (cf. Petit et al. 2011) and sizes in excess of 100 km (cf. Vilenius et al. 2012). Even The Large Synoptic Survey Telescope (LSST) will have a limiting magnitude near this range ($m_R$~24.5; LSST Science Book V2. 2009, p. 18). A far-infrared (FIR) mission with survey capabilities, like the prospective Cryogenic Aperture Large Infrared Space Telescope Observatory (CALISTO; Goldsmith et al. 2008), offers the potential for the first time of really probing the population of TNOs down to moderates sizes, and out to distances exceeding 100 AU from the Sun.

Orbital Resonances with Neptune pump up inclination in the KBOs. Beyond 100AU, the TNO population may flare out as well, with a larger dispersion in inclination, and an increase in the surface density of objects (Morbidelli & Brown 2004). The green curves in Figure 1 (labeled with value of $m_R$) give an idea of the flux density produced by TNOs which are at the limit of detectability at optical wavelengths. CALISTO evidently can go more than one order of magnitude below this, even with predicted confusion limits, indicating the advantage of high sensitivity submillimeter observations of TNO thermal emission, and may go fainter with repeated observations of the field when the object has moved off of background sources.

The ability to derive large quantities of size measurements is a unique value of such FIR surveys. Small bodies typically can vary in their surface reflectivity by factors of 5 or more, while surveys that detect emitted light provide reliable sizes from the flux (cf. Mainzer et al. 2011). This is important because the previous optical surveys have provided alternate size frequency distributions, based on inferences of reflectivity, indicative of competing evolution histories for these bodies (Trujillo et al. 2001; Bernstein et al. 2004; Schlichting et al. 2013), especially at the smaller end (TNO diameters <100 km) of the size scales. Objects at TNO distances will be best detected at wavelengths near 110 $\mu$m. Shorter (~50 $\mu$m), and longer (~200 $\mu$m), wavelengths will better constrain the sizes and temperatures of the objects observed.

*Expected Populations:* Presently, most surveys have placed order-of-magnitude constraints on larger TNOs, with solar-system absolute magnitudes (H)~9 and sizes ~100 km. Petit et al. (2011) place the total of all TNOs, mostly in the classical KBO population near the ecliptic plane, over 100 km in size at ~130,000 in number, and Scattered Disk Objects (SDOs) down to similar sizes, more widely distributed in orbital inclination, near 25,000 in number. Schlichting et al. (2013) and Trujillo et al. (2001) place a cumulative size frequency distribution exponent value q~4, where the number of bodies N with diameters > D go as:

$$N(> D) \propto D^{1-q}$$

so that if, as Figure 1 suggests, a CALISTO-type survey of 1/10$^{th}$ of the sky is sensitive down to TNO diameters D~50 km or smaller, such a survey may yield several tens of thousands of new TNO discoveries, and a correspondingly large sample of TNO sizes, as well as thousands of new SDOs and diameters.

*Related Activity in Related Populations:* CALISTO also has the potential for detecting the limits of cometary activity in these and related populations. Species such as CO may drive sublimation out to distances of several tens of AU (Meech and Svoren, 2004). Detection of extended moving objects within a field owing to the presence of gas and dust coma is possible, and the expected size of such features would extend over several beam widths (A. J. Lovell, private communication). Such observations would place key constraints on the rates of mass lost to ejection of dust from these bodies, as well as the abundance of rarely-observed extremely-volatile species that may be relatively depleted in short-period comets (cf. Bauer et al. 2011, 2012). The onset of such distant activity may be linked to observational phenomenon heretofore unexplained, such as the source of the Centaur color bimodality (the red and gray sub-populations; Tegler et al. 2008), as well as place constraints on the primordial conditions under which they were formed.